\documentclass[pdflatex,sn-mathphys-num]{sn-jnl}

\usepackage{graphicx}%
\usepackage{multirow}%
\usepackage{amsmath,amssymb,amsfonts}%
\usepackage{amsthm}%
\usepackage{mathrsfs}%
\usepackage[title]{appendix}%
\usepackage{xcolor}%
\usepackage{textcomp}%
\usepackage{manyfoot}%
\usepackage{booktabs}%
\usepackage{algorithm}%
\usepackage{algorithmicx}%
\usepackage{algpseudocode}%
\usepackage{listings}%
\usepackage{mathrsfs}%
\usepackage{float}%
\usepackage{subcaption}%

\theoremstyle{thmstyleone}%
\theoremstyle{thmstyletwo}%

\theoremstyle{thmstylethree}%

\begin{document}

\title[Article Title]{EnzyCLIP: A Cross-Attention Dual Encoder Framework 
with Contrastive Learning for Predicting Enzyme Kinetic 
Constants}


\author[1]{\fnm{Anas Aziz} \sur{Khan}}
\email{anasazizkhan555@gmail.com}

\author*[2]{\fnm{Md. Shah} \sur{Fahad}}
\email{fahad8siddiqui@bitmesra.ac.in}

\author[3]{\fnm{Priyanka} \sur{}}
\email{reachpriyanka20@gmail.com}

\author[3]{\fnm{Ramesh} \sur{Chandra}}
\email{rameshchandra@bitmesra.ac.in}

\author[2]{\fnm{Guransh } \sur{Singh}}
\email{guransh766@gmail.com}

\affil[1]{\orgdiv{SCOPE}, \orgname{Vellore Institute of Technology, Vellore}, 
\orgaddress{\street{Katpadi}, \city{Vellore}, \postcode{632014}, \state{Tamil Nadu}, \country{India}}}

\affil*[2]{\orgdiv{Department of Computer Science}, \orgname{BIT}, 
\orgaddress{\street{Mesra}, \city{Ranchi}, \postcode{835215}, \state{Jharkhand}, \country{India}}}

\affil[3]{\orgdiv{Department of Bioengineering and Biotechnology}, \orgname{BIT}, 
\orgaddress{\street{Mesra}, \city{Ranchi}, \postcode{835215}, \state{Jharkhand}, \country{India}}}


\abstract{
Accurate prediction of enzyme kinetic parameters is crucial for drug discovery, metabolic engineering, and synthetic biology applications. Current computational approaches face limitations in capturing complex enzyme–substrate interactions and often focus on single parameters while neglecting the joint prediction of catalytic turnover numbers (K$_{cat}$) and Michaelis–Menten constants (K$_m$). We present {EnzyCLIP}, a novel dual-encoder framework that leverages contrastive learning and cross-attention mechanisms to predict enzyme kinetic parameters from protein sequences and substrate molecular structures. Our approach integrates {ESM-2} protein language model embeddings with {ChemBERTa} chemical representations through a CLIP-inspired architecture enhanced with bidirectional cross-attention for dynamic enzyme–substrate interaction modeling. EnzyCLIP combines {InfoNCE contrastive loss} with {Huber regression loss} to learn aligned multimodal representations while predicting log$_{10}$-transformed kinetic parameters. EnzyCLIP is trained on the CatPred-DB database containing 23,151 K$_{cat}$ and 41,174 K$_m$ experimentally validated measurements, and achieved competitive baseline performance with {R$^2$ scores of 0.593 for K$_{cat}$ and 0.607 for K$_m$ prediction}. XGBoost ensemble methods on learned embeddings further improved {K$_m$ prediction (R$^2$ = 0.61)} while maintaining robust K$_{cat}$ performance.
}

\maketitle


\section{Introduction}\label{sec1}

Enzymes are catalysts that enable reactions to happen at rapid rates under the mild conditions of temperature, pH, and pressure inside cells. In contrast to inorganic catalysts, which mostly require harsh or extreme conditions to function, enzymes can work powerfully under gentle conditions-namely, near neutral pH, moderate temperatures, and in aqueous environments. This is a reflection of how evolution has precisely honed them to sustain the fragile balance of cellular life\cite{ref1,ref3}.

Essentially, enzymes are involved in almost all cellular processes. They perform major metabolic routes, such as glycolysis, the tricarboxylic acid cycle, and oxidative phosphorylation-which jointly supply energy for the cell-apart from metabolism, replication of DNA and repair, transcription, translation of proteins, neurotransmitter breakdown, and cell signaling are some other areas in which enzymes have important roles. Given such a wide array of their functions, there is no doubt that they are very much part of cellular health and the amazing complexity of life\cite{ref6}.

Predictions of protein-ligand binding free energy are difficult to interpret unless reliable estimates of experimental uncertainties in measured binding free energies are provided.

The extraordinary catalytic efficiency of enzymes comes from their capability for lowering the height of the activation energy barrier that slows down chemical transformations\cite{ref2}. Enzymes, through stabilizing transition states and orienting reactive groups in precise spatial arrangements, make it energetically easier for reactants to convert into products. Usually, each enzyme is highly specific towards certain molecular partners called substrates. This comes from structural complementarity between the enzyme active site and the geometry of the substrate molecule\cite{ref4}.

The active site is a pocket in a protein, comprising only 10\% to 20\% of the total volume of the enzyme. It is shaped when amino acids from different parts fold together. Binding of the substrate in this site by weak non-covalent interactions such as hydrogen bonds, electrostatic attractions, and van der waals forces develops a temporary enzyme-substrate complex where the chemical reaction occurs. The enzyme will return to its original conformation after the reaction is complete and can begin processing another substrate molecule\cite{ref5}.

With the development of enzyme kinetics, a field that measures and models the rates of enzymatic reactions, scientists are able to evaluate how efficiently enzymes carry out their catalytic functions. Lying at the heart of the study of enzyme kinetics is the Michaelis-Menten model, which describes how the rate of reaction varies with substrate concentration. The model is mathematically expressed as:

\begin{equation}
v = \frac{V_{\max} \cdot [S]}{K_m + [S]}
\end{equation}

where $v$ is the reaction velocity, $[S]$ is the substrate concentration, $V_{\max}$ is the maximum achievable rate of reaction, and $K_m$ is the Michaelis constant\cite{ref7}.

The model helps in determining two key parameters: the turnover number ($K_{\text{cat}}$), representing substrate molecules converted per enzyme molecule per unit time at maximum efficiency, and the Michaelis constant ($K_m$), denoting substrate concentration at half-maximal reaction rate. The relationship between $V_{\max}$ and $K_{\text{cat}}$ is given by:

\begin{equation}
K_{\text{cat}} = \frac{V_{\max}}{[E]_{\text{total}}}
\end{equation}

where $[E]_{\text{total}}$ is the total enzyme concentration. The catalytic efficiency, expressed as $K_{\text{cat}}/K_m$, serves as a comprehensive metric for enzyme performance under physiological conditions where substrate saturation rarely occurs\cite{ref8}.

Accurate prediction of kinetic parameters such as $K_{\text{cat}}$ and $K_m$ helps to advance various fields including drug discovery, metabolic engineering, and synthetic biology. For instance, during drug design, insight into the way different kinds of compounds are being processed by enzymes could inform strategies in the creation of molecules possessing favorable pharmacokinetic properties. In metabolic engineering, knowledge of such parameters enables the fine-tuning of enzyme activity and expression levels with the aim of improving the efficiency of biochemical production pathways\cite{ref9}. However, determination of these values experimentally is mostly laborious and expensive, and the process cannot easily be scaled to cover the huge diversity of possible enzyme-substrate interactions.

Historically, the computational prediction of enzyme kinetics had been done by rule-based systems built on expert knowledge, QSAR models connecting molecular features to biological activity, and molecular docking for estimating binding strength. However, most of these traditional approaches relied much on hand-crafted features and expert knowledge. Hence, they were difficult to scale up and less effective at handling new or unseen enzyme-substrate pairs.

The emergence of machine learning and deep learning has transformed the prediction of enzyme kinetics. Contemporary models employ natural language processing techniques for biological data, regarding a protein sequence as a sentence comprising amino acids while representing substrates as SMILES strings\cite{ref11}. Pretrained models learn complex contextual relationships between the elements of the sequence in large-scale training with unlabelled data. After fine-tuning for the kinetic parameters prediction, they often outperform traditional methods not only for their accuracy but also for generalization. Deep learning frameworks can incorporate several data modalities; they combine protein sequence embeddings with molecular graph representations for creating an integrated representation of enzyme-substrate pairs.

Some recent works include DeepEnzyme\cite{ref12}, which integrates Transformer and Graph Convolutional Networks to model both the sequence and 3D structural information in order to better predict $K_{\text{cat}}$, MPEK\cite{ref45}, a multitask model that can perform parallel prediction of $K_{\text{cat}}$ and $K_m$ by taking environmental factors into account\cite{ref13}, and CatPred, a unified framework for integrating sequence-attention mechanisms with equivalent graph neural networks to predict kinetic parameters for protein sequences\cite{ref43}. These computational methods are further accelerating enzyme discovery and engineering, enabling the high-throughput screening of large datasets and better information for biological system modeling.

In this work, we present a new CLIP-inspired dual-encoder framework, EnzyCLIP, which is designed for enzyme-substrate kinetic predictions. The model focuses on predicting two main kinetic parameters: $K_{\text{cat}}$ (turnover number) and $K_m$ (Michaelis constant). EnzyCLIP integrates deep learning with contrastive representation learning to jointly embed protein sequences and small-molecule substrates within a shared latent space, which allows the model to effectively capture and quantify functional relationships between enzymes and their substrates.

We utilize a dual encoder architecture wherein two separate biological modalities, protein sequences and substrate chemical structures, are independently processed. Each modality is fed into a different transformer-based encoder that generates fixed dimensional embeddings capturing the key biological and chemical features, respectively. The model leverages contrastive learning to be trained on associating the correct enzyme-substrate pairs by maximizing the similarity between their embeddings while minimizing the similarity for incorrect or random combinations. In this learning framework, subtle patterns that define substrate specificity and catalytic compatibility can be learned.

Specifically, EnzyCLIP uses ESM-2, a deep protein language model that is trained on hundreds of millions of amino acid sequences to extract rich contextual embeddings from enzymes. ESM-2 captures evolutionarily and structurally meaningful features of proteins within a high-dimensional latent space. These embeddings act as biologically meaningful features that help in downstream functional prediction \cite{ref15}.

In parallel, substrate molecules are encoded by ChemBERTa, which is a BERT-based transformer trained on the SMILES representation of chemical compounds. ChemBERTa learns molecular structure and function in a language-like fashion. It learns to recognize patterns of chemical reactivity and substructure directly from tokenized molecular strings \cite{ref16}. EnzyCLIP maps enzyme and substrate embeddings to a shared latent space with the help of contrastive learning by utilizing learned projection layers \cite{ref17}. In the process, it aligns the correct pairs while separating the mismatched pairs. The architecture also integrated bi-directional cross-attention \cite{ref18} since there are dynamic protein-substrate interactions, and it predicts using Huber Loss \cite{ref19} $\log_{10}$-transformed $K_{\text{cat}}^{\max}$ and $\log_{10} K_m$ values for better robustness. Unlike traditional approaches, EnzyCLIP learns directly from raw sequences, without relying on handcrafted features, offering scalability and interpretability for downstream tasks such as enzyme engineering and activity screening. This unified framework thus marks a quantum jump forward in AI-driven enzymology.

\section{Literature Review}\label{sec2}

Traditional methods for determining kinetic parameters experimentally are time-consuming, expensive, and not suitable for large-scale studies. Thousands of enzymes remain uncharacterized with the rapid growth of metagenomic data. This has created a strong demand for computational methods that can analyze enzymes and predict their functions on a large scale in a much faster way.

The early computational approaches are highly dependent on manually crafted features and domain expert knowledge. QSAR models utilize predefined molecular descriptors for relating activity to the molecular structure with poor transferability and require extensive feature engineering\cite{ref20}. Homology modeling combined with molecular docking provided useful mechanistic insights but was computationally intensive and required the availability of accurate structural data\cite{ref21}. Traditional machine learning methods utilize amino acid composition and physiochemical properties as input features, such as Support Vector Regression (SVR)\cite{ref32}, Random Forests\cite{ref23}, and Decision Trees\cite{ref24}. These methods generalized poorly for new, uncharacterized enzyme families\cite{ref25} and remained dependent on considerable domain-specific manual feature design\cite{ref26}. In general, these classical methods had the following critical limitations: reliance on handcrafted features, use of 3D structural information, and poor scalability across diverse enzyme classes \cite{ref27}

The development of machine learning applications for enzyme kinetics prediction has been greatly enhanced by curated databases such as BRENDA\cite{ref28} and SABIO-RK\cite{ref29}. Deep learning eliminated the need to rely on hand-crafted feature engineering, as it amply captures complex, non-linear relationships involving biological systems. Some protein language models, such as ESM-1b/2\cite{ref41}, learn dense embeddings that encode structural, functional, and evolutionary information via self-supervised training. Chemical language models, such as ChemBERTa\cite{ref16}, MolFormer\cite{ref36}, and GROVER\cite{ref42}, offer state-of-the-art molecular representation learned from SMILES strings directly. All of these collectively offer the possibility of end-to-end learning wherein a model will be able to discover meaningful features automatically, generate context-aware representations, and make full use of transfer learning across diverse biochemical tasks. More specialized frameworks include DLKcat\cite{ref40}, CatPred\cite{ref43}, and MPEK\cite{ref45}, which achieved marked improvement compared to the traditional methods. Graph Convolutional Networks enhance prediction accuracy by integrating sequence-derived insights about evolution with structure-based geometric features\cite{ref46}.

CLIP-inspired dual encoder frameworks learn shared representations across biological modalities by separately embedding enzymes and substrates and aligning them through contrastive learning objectives\cite{ref47}. The contrastive InfoNCE loss function drives alignment:

\begin{equation}
L_{\text{contrastive}} = -\log \frac{\exp(\text{sim}(e_i, s_i)/\tau)}{\sum_{j=1}^N \exp(\text{sim}(e_i, s_j)/\tau)}
\end{equation}

where $\text{sim}(e_i, s_i)$ indicates the similarity between enzyme embedding $e_i$ and substrate embedding $s_i$, $\tau$ is the temperature parameter, and $N$ is the batch size. This will allow the model to capture biochemical relationships such as substrate preferences and cofactor requirements that it can use to make predictions even for enzyme–substrate pairs it has never seen \cite{ref48}.

Cross-attention mechanisms allow for a two-way interaction between enzyme and substrate representations, modelling biological phenomena such as induced fit and allosteric regulation. Multi-task learning frameworks jointly predict both parameters:

\begin{equation}
L_{\text{total}} = \alpha L_{\text{contrastive}} + \beta L_{K_{\text{cat}}} + \gamma L_{K_m}
\end{equation}

where $\alpha$, $\beta$, and $\gamma$ control relative importance of each loss component\cite{ref49}. In contrast to the traditional supervised methods, contrastive frameworks generalise better, are more data efficient and robust.

The current computational methods have a number of key challenges strongly limiting practical usability. Treating enzyme and substrate information in a separated way ignores mutual dependencies and dynamic interactions defining enzymatic catalysis \cite{ref50}. In another important methodological gap, the prevailing focus is on $K_{\text{cat}}$, while $K_m$ is usually neglected, though both parameters emerge from the same catalytic dynamics. A number of the existing models behave like black boxes, featuring limited biological interpretability and controllable molecular factors driving catalytic efficiency. This issue restricts their usability for enzyme engineering. Finally, poor generalization to underrepresented families introduces major problems in the case of metagenomic enzyme discovery \cite{ref51}. Enzymatic catalysis is inherently a dynamic process. Substrate binding induces changes in enzyme shape, such as an induced fit mechanism, allosteric effects, or specificity depending on the molecular context. However, such complex behaviors can hardly be modeled when enzyme and substrate information are processed separately. Both $K_{\text{cat}}$ and $K_m$ depend on interdependent factors involving binding strength, structural events, chemical transformations, and product release. Such challenges provide rationales for further development that will allow the integration of contrastive learning with cross-attention mechanisms into models capable of jointly predicting kinetic parameters and delivering much clearer and more interpretable insight into how enzymes work.

\section{Methodology}

This section details the experimental setting and technical framework of EnzyCLIP. We take the CatPed-DB\cite{ref43} dataset containing 23,197 entries for $K_{\text{cat}}$ prediction and 41,174 entries for $K_m$ prediction. Our proposed method combines careful dataset curation with state-of-the-art molecular representation learning, contrastive learning strategy, and deep neural network architecture. Enzyme amino acid sequences are represented using ESM-2\cite{ref41} transformer model, while substrate molecules are encoded using ChemBERTa\cite{ref16}, thus allowing deep and biologically meaningful representation of both modalities. After processing, embeddings are fed into a CLIP-inspired dual encoder enhanced by a bi-directional cross-attention mechanism. The loss function combines InfoNCE contrastive loss\cite{ref52} to enforce semantic closeness between biochemically compatible enzyme-substrate pairs and incorporates Smooth L1 loss\cite{ref19} for quantitative prediction of kinetic parameters.  

We implemented dataset preprocessing using \texttt{pandas} DataFrame operations\cite{ref53} and applied quality control measures in a systematic way. Regarding the $K_m$ estimation, we first loaded from the CSV file and then filtered it to retain only entries where information in all the three columns \texttt{sequence}, \texttt{substrate\_smiles}, and \texttt{log\textsubscript{10}km\_mean} was not missing. We removed any rows with NaN values or whitespace-only strings. To predict $K_{\text{cat}}$, we followed the same process and filtered the dataset to include only records where valid values were present in the columns \texttt{sequence}, \texttt{substrate\_smiles}, and \texttt{log\textsubscript{10}kcat\_max}. We excluded any incomplete or missing entries to guarantee the consistency and reliability of the data in both kinetic parameter prediction tasks.

\begin{equation}
\text{Filtered Dataset} = \{(s_i, m_i, y_i) : s_i \neq \emptyset, m_i \neq \emptyset, y_i \neq \text{NaN}\}
\label{eq:dataset_filtering}
\end{equation}

where $s_i$, $m_i$, and $y_i$ denote the protein sequence, substrate SMILES\cite{ref54} string, and the logarithmic kinetic parameter of entry $i$, respectively. The curated dataset was subjected to systematic partition into training, validation, and test sets in an 80:10:10 ratio with a fixed random seed of 42 using the PyTorch \texttt{random\_split} function for reproducibility (Table~\ref{tab:dataset_composition}).

\begin{equation}
\begin{aligned}
\text{train\_size} &= \lfloor 0.8 \times \text{total\_size} \rfloor \\
\text{val\_size} &= \lfloor 0.1 \times \text{total\_size} \rfloor \\
\text{test\_size} &= \text{total\_size} - \text{train\_size} - \text{val\_size}
\end{aligned}
\label{eq:data_split}
\end{equation}

\begin{table}[h]
\caption{Dataset composition after preprocessing and splitting procedures}
\label{tab:dataset_composition}
\begin{tabular*}{\textwidth}{@{\extracolsep\fill}lcccc}
\toprule
Parameter & Training Set (80\%) & Validation Set (10\%) & Test Set (10\%) & Total \\
\midrule
$K_m$ Dataset Entries & Variable\footnotemark[1] & Variable\footnotemark[1] & Variable\footnotemark[1] & Filtered Total \\
Split Ratio & 0.8 & 0.1 & 0.1 & 1.0 \\
Random Seed & 42 & 42 & 42 & 42 \\
Batch Size & 64 & 64 & 64 & 64 \\
\botrule
\end{tabular*}
\footnotetext[1]{Exact numbers depend on filtering results and are computed dynamically during preprocessing.}
\footnotetext{Note: The $K_{\text{cat}}$ dataset follows identical preprocessing procedures with 23,197 total entries, while $K_m$ dataset contains 41,174 entries from CatPred-DB.}
\end{table}

The molecular representation strategy employed pre-trained language models specifically designed for biological and chemical sequences. Enzyme sequences were processed using the ESM-2\cite{ref41} transformer architecture, specifically the \texttt{facebook/esm2\_t6\_8M\_UR50D} variant with 8 million parameters. This model learns evolutionary, structural, and functional patterns through extensive pretraining on multiple sequence alignment databases. Substrate molecules were encoded using the ChemBERTa\cite{ref16} \texttt{seyonec/ChemBERTa-zinc-base-v1} model, pretrained on the ZINC database. Both tokenizers utilized maximum sequence length of 256 tokens with \texttt{max\_length} padding and truncation enabled.

\begin{equation}
\mathbf{T}_{\text{prot}} = \text{ESM2Tokenizer}(s;\ \text{max length}=256)
\label{eq:protein_tokenization}
\end{equation}

\begin{equation}
\mathbf{T}_{\text{chem}} = \text{ChemBERTaTokenizer}(m;\ \text{max length}=256)
\label{eq:chemical_tokenization}
\end{equation}

where, $s$ denotes the protein sequence and $m$ denotes the substrate SMILES string.We implemented the class \texttt{EnzyCLIPDataset} as a subclass of PyTorch's Dataset that handled the complete preprocessing and tokenization pipeline. It tokenized on-the-fly in the \texttt{\_\_getitem\_\_} method for memory efficiency; thus, it returned \texttt{input\_ids} and \texttt{attention\_mask} tensors for both protein and chemical modalities. Finally, we created for all the sets DataLoader objects with batch size 128;, shuffling was enabled only for training sets.

The core neural architecture integrates contrastive learning principles with cross-attention mechanisms, together with regression prediction capabilities. The architecture makes use of frozen pre-trained encoders for both protein and chemical modalities, maintaining valuable representations while concentrating optimization on task-specific components. Each encoder is followed by projection heads implemented as nn.Sequential modules mapping outputs to a common 256-dimensional embedding space.

\begin{equation}
\mathbf{P}_{\text{prot}} = \text{LayerNorm}(\mathbf{W}_p \mathbf{h}_{\text{prot}} + \mathbf{b}_p)
\label{eq:protein_projection}
\end{equation}

\begin{equation}
\mathbf{P}_{\text{chem}} = \text{LayerNorm}(\mathbf{W}_c \mathbf{h}_{\text{chem}} + \mathbf{b}_c)
\label{eq:chemical_projection}
\end{equation}

where, $\mathbf{h}_{\text{prot}}$ and $\mathbf{h}_{\text{chem}}$ denote the protein and chemical encoder outputs, respectively.This architecture integrates bi-directional cross-attention mechanisms implemented through \texttt{nn.MultiheadAttention} with four attention heads that enable explicit modeling of conditional dependencies between enzyme and substrate representations. Such a design avoids the limitation of simple concatenation, because it allows the model to selectively focus on the most relevant features across both modalities, hence leading to richer and more context-aware representations.

\begin{equation}
\mathbf{F}_{\text{prot}} = \text{CrossAttention}(\mathbf{P}_{\text{prot}}^{\text{unsq}},\ \mathbf{P}_{\text{chem}}^{\text{unsq}},\ \mathbf{P}_{\text{chem}}^{\text{unsq}})
\label{eq:cross_attention_protein}
\end{equation}

\begin{equation}
\mathbf{F}_{\text{chem}} = \text{CrossAttention}(\mathbf{P}_{\text{chem}}^{\text{unsq}},\ \mathbf{P}_{\text{prot}}^{\text{unsq}},\ \mathbf{P}_{\text{prot}}^{\text{unsq}})
\label{eq:cross_attention_chemical}
\end{equation}

where, $\mathbf{P}_{\text{prot}}^{\text{unsq}}$ and $\mathbf{P}_{\text{chem}}^{\text{unsq}}$ represent the unsqueezed projected embeddings used as query, key, and value inputs in the cross-attention mechanism.The regression head converts the concatenated cross-attention outputs using a three-layer feedforward network with dimensions of 256 → 128 → 1. It uses GELU activations to introduce non-linearity and applies dropout regularization at rates of 0.2 and 0.1 to avoid overfitting (Table~\ref{tab:architecture_specs}).

\begin{equation}
\mathbf{h}_{\text{concat}} = \text{Concat}[\mathbf{F}_{\text{prot}}, \mathbf{F}_{\text{chem}}]
\label{eq:concatenation}
\end{equation}

\begin{equation}
\hat{y} = \mathbf{W}_3(\text{GELU}(\text{Dropout}(\mathbf{W}_2(\text{GELU}(\text{Dropout}(\mathbf{W}_1 \mathbf{h}_{\text{concat}} + \mathbf{b}_1))) + \mathbf{b}_2))) + \mathbf{b}_3
\label{eq:regression_network}
\end{equation}

Here, $\mathbf{h}_{\text{concat}}$ denotes the fused embedding, $\hat{y}$ is the predicted kinetic parameter, and $\mathbf{b}_1$, $\mathbf{b}_2$, $\mathbf{b}_3$ are the bias vectors in the regression layers.

\begin{table}[h]
\caption{Neural architecture specifications as implemented in the Model}
\label{tab:architecture_specs}
\begin{tabular*}{\textwidth}{@{\extracolsep\fill}lc}
\toprule
Component & Specification \\
\midrule
Protein Encoder & facebook/esm2\_t6\_8M\_UR50D (frozen) \\
Chemical Encoder & seyonec/ChemBERTa-zinc-base-v1 (frozen) \\
Shared Embedding Dimension & 256 \\
Cross-Attention Heads & 4 \\
Regression Layer 1 & 256 → 128 \\
Regression Layer 2 & 128 → 1 \\
Activation Function & GELU \\
Dropout Rate (Layer 1) & 0.2 \\
Dropout Rate (Layer 2) & 0.1 \\
Temperature Parameter (Initial) & 0.07 (learnable) \\
Normalization & Layer Normalization \\
\botrule
\end{tabular*}

\end{table}

The training framework follows a multi-objective optimization strategy that jointly balances contrastive representation learning with supervised learning via supervised regression. The contrastive learning component utilizes a symmetric InfoNCE loss\cite{ref52}, which assumes batch-wise correspondence between modalities, where each index corresponds to an experimentally validated enzyme-substrate pair. For computing the similarity we first perform L2 normalization of the projected embeddings followed by matrix multiplication scaled by a learnable temperature parameter.

\begin{equation}
\mathbf{L}_{\text{sim}} = \frac{\mathbf{Z}_{\text{prot}} \mathbf{Z}_{\text{chem}}^T}{\exp(\tau)}
\label{eq:similarity_logits}
\end{equation}

This symmetric InfoNCE loss calculates cross-entropy on both the similarity matrix and its transpose, ensuring that the contrastive objectives are bidirectional. Here, $\mathbf{Z}_{\text{prot}}$ and $\mathbf{Z}_{\text{chem}}$ represent the $\ell_2$-normalized protein and chemical embeddings, respectively, and $\tau$ represents the learnable temperature parameter.

\begin{equation}
\mathcal{L}_{\text{InfoNCE}} = \frac{1}{2}[\mathcal{L}_{\text{CE}}(\mathbf{L}_{\text{sim}}, \mathbf{y}) + \mathcal{L}_{\text{CE}}(\mathbf{L}_{\text{sim}}^T, \mathbf{y})]
\label{eq:symmetric_infonce}
\end{equation}

The regression loss uses \texttt{nn.SmoothL1Loss} (Huber loss)\cite{ref19} for robustness to experimental outliers, combining quadratic penalties for small errors with linear penalties for large deviations. Here, $\mathbf{y}$ refers to the ground-truth index targets for contrastive matching across protein–chemical pairs.

\begin{equation}
\mathcal{L}_{\text{SmoothL1}}(\hat{y}, y) = \begin{cases}
0.5(\hat{y} - y)^2 & \text{if } |\hat{y} - y| < 1 \\
|\hat{y} - y| - 0.5 & \text{otherwise}
\end{cases}
\label{eq:smooth_l1_loss}
\end{equation}

The total training loss combines both objectives in an equal weighting manner to ensure the balanced optimization of representation learning and predictive accuracy.

\begin{equation}
\mathcal{L}_{\text{total}} = \mathcal{L}_{\text{InfoNCE}} + \mathcal{L}_{\text{SmoothL1}}
\label{eq:total_loss}
\end{equation}

We trained the model by using AdamW\cite{ref55} as an optimizer with an initial learning rate of $2 \times 10^{-4}$ and weight decay of $1 \times 10^{-2}$. The learning rate was adjusted during training by using a cosine annealing schedule ($T_{\max} = 20$) over 25 epochs. To improve computational efficiency and stabilize the training, we enabled Automatic Mixed Precision (AMP) and performed gradient clipping with a maximum norm of 1.0 to avoid gradient explosion. By default, model checkpoints are saved according to the best validation $R^2$ score.

\begin{equation}
\eta_t = \eta_{\min} + \frac{1}{2}(\eta_{\max} - \eta_{\min})\left(1 + \cos\left(\frac{t}{T_{\max}}\pi\right)\right)
\label{eq:cosine_annealing}
\end{equation}

Advanced training techniques included Automatic Mixed Precision (AMP) using torch.amp.autocast and GradScaler, which allows half-precision computation for efficiency. Gradient clipping with maximum norm 1.0 to avoid gradient explosion\cite{ref56}.

\begin{equation}
\mathbf{g}_{\text{clipped}} = \mathbf{g} \cdot \min\left(1, \frac{1.0}{||\mathbf{g}||_2}\right)
\label{eq:gradient_clipping}
\end{equation}

Comprehensive regression metrics were used for model evaluation, including Mean Squared Error (MSE), Root Mean Squared Error (RMSE), Mean Absolute Error (MAE), Coefficient of Determination ($R^2$), and Pearson's correlation coefficient ($r$). Besides, residual-based statistics like mean and standard deviation of residuals were calculated to further check the model for calibration and bias. These metrics collectively quantify both predictive accuracy and consistency across validation and test sets.

\begin{equation}
\text{MSE} = \frac{1}{n} \sum_{i=1}^{n} (y_i - \hat{y}_i)^2, \quad
\text{RMSE} = \sqrt{\frac{1}{n} \sum_{i=1}^{n} (y_i - \hat{y}_i)^2}
\label{eq:mse_rmse}
\end{equation}

\begin{equation}
\text{MAE} = \frac{1}{n} \sum_{i=1}^{n} |y_i - \hat{y}_i|, \quad
R^2 = 1 - \frac{\sum_{i=1}^{n} (y_i - \hat{y}_i)^2}{\sum_{i=1}^{n} (y_i - \bar{y})^2}
\label{eq:mae_r2}
\end{equation}

\begin{equation}
r = \frac{\sum_{i=1}^{n}(y_i - \bar{y})(\hat{y}_i - \bar{\hat{y}})}{\sqrt{\sum_{i=1}^{n}(y_i - \bar{y})^2} \sqrt{\sum_{i=1}^{n}(\hat{y}_i - \bar{\hat{y}})^2}}
\label{eq:pearson}
\end{equation}

where, $y_i$ and $\hat{y}_i$ denote the ground-truth and predicted kinetic parameters, $\bar{y}$ and $\bar{\hat{y}}$ are their respective means, and $n$ is the number of samples.

In our evaluation, we extracted both predictions and immediate features embeddings from the projection layers across all dataset splits. Each sample generated protein- and substrate-derived latent representations, denoted as $\mathbf{e}_{\text{prot}}$ and $\mathbf{e}_{\text{chem}}$, respectively. We avoided the typical CLS-token summarization and instead used mean pooling on token embeddings, weighted by attention masks, to get fixed-length representations for each modality. This approach ensured that only contextually relevant tokens contribute proportionally to the final molecular embeddings, therefore enhancing the robustness and expensiveness of the learned representation space.

After model convergence, multimodal embeddings were extracted from the projection layers for all dataset splits to enable downstream regression analysis. Each sample produced both enzyme-derived and substrate-derived latent representations, denoted $\mathbf{e}_{\text{prot}}$ and $\mathbf{e}_{\text{chem}}$, respectively. We applied mean pooling over token-level hidden states, which is weighted by attention masks in order to generate fixed-length contextual representations. This ensured that only valid tokens contribute proportionally to the final embedding:

\begin{equation}
\mathbf{e} = \frac{\sum_{t=1}^{L} h_t \cdot a_t}{\sum_{t=1}^{L} a_t}
\label{eq:mean_pooling}
\end{equation}

 where, $h_t$ denotes the hidden state at position $t$, $a_t$ is the corresponding attention-mask weight, and $L$ is the sequence length. The resulting mean-pooled vectors were then passed through modality-specific projection layers to obtain compact latent representations in a shared 256-dimensional embedding space:

\begin{equation}
\mathbf{e}_{\text{prot}} = \text{Proj}_{\text{prot}}(\mathbf{e}_{\text{mean,prot}}), \quad
\mathbf{e}_{\text{chem}} = \text{Proj}_{\text{chem}}(\mathbf{e}_{\text{mean,chem}})
\label{eq:projected_embeddings}
\end{equation}

The projected representations were then concatenated to form a unified embedding vector that jointly captures the enzyme–substrate interaction context:

\begin{equation}
\mathbf{e}_{\text{combined}} = \text{Concat}[\mathbf{e}_{\text{prot}}, \mathbf{e}_{\text{chem}}]
\label{eq:combined_embedding}
\end{equation}

The extracted representations were subsequently used as high-level molecular embeddings in downstream regression through ensemble-based learning frameworks. In this work, two gradient-boosting regressors, XGBoost\cite{ref57} and CatBoost\cite{ref58}, were fitted on the extracted features to model the nonlinear mapping between latent biochemical representations and kinetic parameters. The XGBoost was set with 500 estimators, maximum depth of 10, learning rate of 0.05, and subsampling ratio of 0.8, while CatBoost regressor utilized 500 iterations with similar depth and learning rate, optimized with RMSE loss. Both regressors were evaluated using the same training–validation–test splits of the EnzyCLIP model for consistent benchmarking across the different prediction pipelines.

\begin{equation}
\hat{y}_{\text{XGB}} = \sum_{k=1}^{500} f_k(\mathbf{e}_{\text{combined}}), \quad
\hat{y}_{\text{CB}} = \sum_{k=1}^{500} g_k(\mathbf{e}_{\text{combined}})
\label{eq:ensemble_regressors}
\end{equation}

This embedding-based ensemble regression step assessed the generalizability and expressiveness of the representations learned by EnzyCLIP. The methodology thus integrates deep contrastive feature extraction with classical ensemble regression techniques to provide a comprehensive and interpretable modeling framework for enzyme-substrate kinetic parameter prediction.

\section{Results}
EnzyCLIP learns robust, generalized multimodal representations across the kinetic parameters $K_{\text{cat}}$ and $K_m$ on diverse enzyme--substrate pairs. The model exhibits competitive test performances for $K_{\text{cat}}$ ($R^2 = 0.60$) and $K_m$ ($R^2 = 0.61$), outperforming traditional baselines such as Random Forest, XGBoost, SVR, and CatBoost when trained directly on raw inputs. Ensemble regressors trained on EnzyCLIP embeddings further improve prediction accuracy, achieving up to $R^2 \approx 0.61$ for $K_m$, thereby validating the quality of the learned feature space. Stratified analyses show that while $K_m$ predictions remain stable across a wide range of sequence lengths, accuracy in $K_{\text{cat}}$ predictions degrades rapidly for very long proteins ($>800$ aa), underscoring distinct biochemical determinants of catalysis versus binding affinity. EC-class--wise comparisons also highlight variation in predictive difficulty aligned with enzymatic function. Explainability analyses demonstrate that different embedding dimensions from EnzyCLIP govern $K_{\text{cat}}$ versus $K_m$, confirming that the model captures meaningful mechanistic differences between catalytic efficiency and substrate affinity. Ablation studies reveal that multimodal integration is critical, with the removal of either protein or substrate information reducing $R^2$ by an average of 30--50\%. Overall, these results position EnzyCLIP as a strong, generalizable, and mechanistically informative framework for enzyme kinetics prediction.

\subsection{Enzyme Turnover Number (\texorpdfstring{$K_{\text{cat}}$}{kcat}) Prediction}

\subsubsection{Dataset Characteristics and Distribution}

\begin{figure}[H]
\centering
\includegraphics[width=0.78\textwidth]{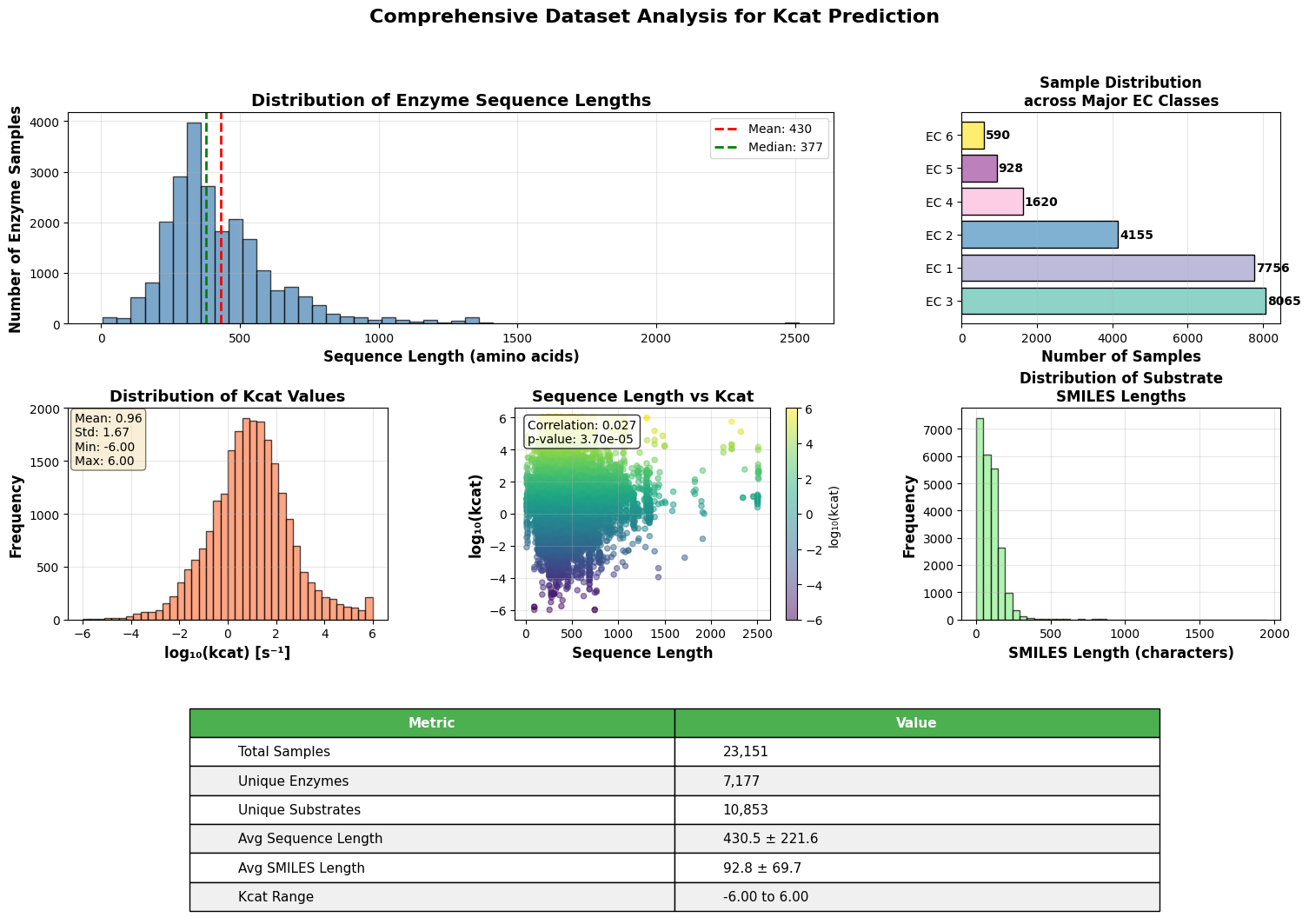}
\caption{Comprehensive \(K_{\text{cat}}\) dataset analysis. Distribution of enzyme sequence lengths showing mean of 430 and median of 377 amino acids (top left), sample distribution across EC classes with EC 3 most represented (top right), \(K_{\text{cat}}\) value distribution with mean 0.96 and range -6 to 6 (bottom left), weak correlation between sequence length and \(K_{\text{cat}}\) (\(r=0.027\), bottom center), and substrate SMILES length distribution (bottom right). Summary statistics table shows 23,151 total samples from 7,177 unique enzymes.}
\label{fig:kcat_comprehensive}
\end{figure}

\begin{figure}[H]
\centering
\includegraphics[width=0.78\textwidth]{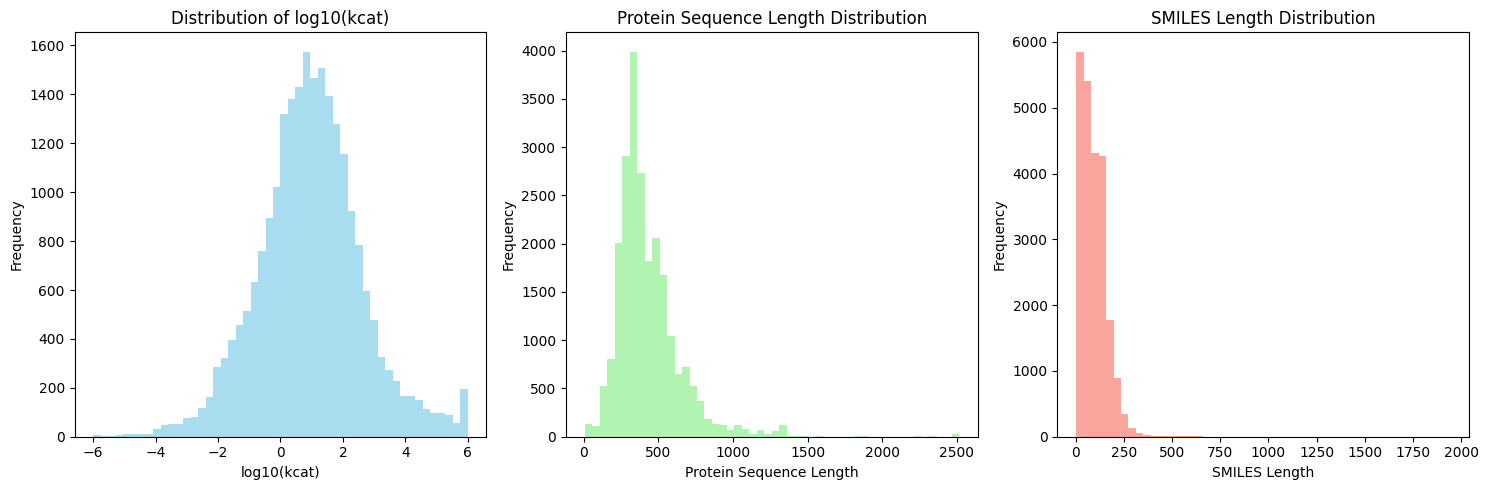}
\caption{Dataset distribution comparisons for \(K_{\text{cat}}\). Three-panel visualization showing (left) \(\log_{10}(K_{\text{cat}})\) distribution with normal characteristics, (center) protein sequence length distribution peaking at 300--500 amino acids, and (right) SMILES length distribution heavily left-skewed with most substrates below 100 characters.}
\label{fig:kcat_dataset_distribution}
\end{figure}

The \(K_{\text{cat}}\) prediction dataset was comprised of 23,151 enzyme-substrate pairs representing 7,177 unique enzymes and 10,853 unique substrates (Fig.~\ref{fig:kcat_comprehensive},Fig.~\ref{fig:kcat_dataset_distribution}). The enzyme sequences had an average length of 430.5 \(\pm\) 221.6 amino acids with a median of 377 amino acids, and substrate SMILES strings averaged 92.8 \(\pm\) 69.7 characters (Fig.~\ref{fig:kcat_comprehensive}). The \(K_{\text{cat}}\) values spanned a 12-order-of-magnitude range from -6.00 to 6.00 (\(\log_{10}\) scale), with a mean of 0.96 and standard deviation of 1.67 (Fig.~\ref{fig:kcat_comprehensive}). There was a weak but statistically significant correlation (\(r = 0.027\), \(p = 3.70 \times 10^{-5}\)) between enzyme sequence length and \(K_{\text{cat}}\) values, suggesting that sequence length has little impact on catalytic turnover (Fig.~\ref{fig:kcat_comprehensive}).

Sample distribution across major Enzyme Commission (EC) classes showed the following representation: with EC 3 (hydrolases) constituting the largest proportion at 8,065 samples, followed by EC 1 (oxidoreductases) with 7,796 samples, EC 2 (transferases) with 4,155 samples, EC 4 (lyases) with 1,620 samples, EC 5 (isomerases) with 928 samples, and finally EC 6 (ligases) with 590 samples (Fig.~\ref{fig:kcat_comprehensive}). This distribution also reflects the natural prevalence of these enzyme classes in CatPred-DB.

\subsubsection{Model Training and Performance}

\begin{figure}[H]
\centering
\includegraphics[width=0.7\textwidth]{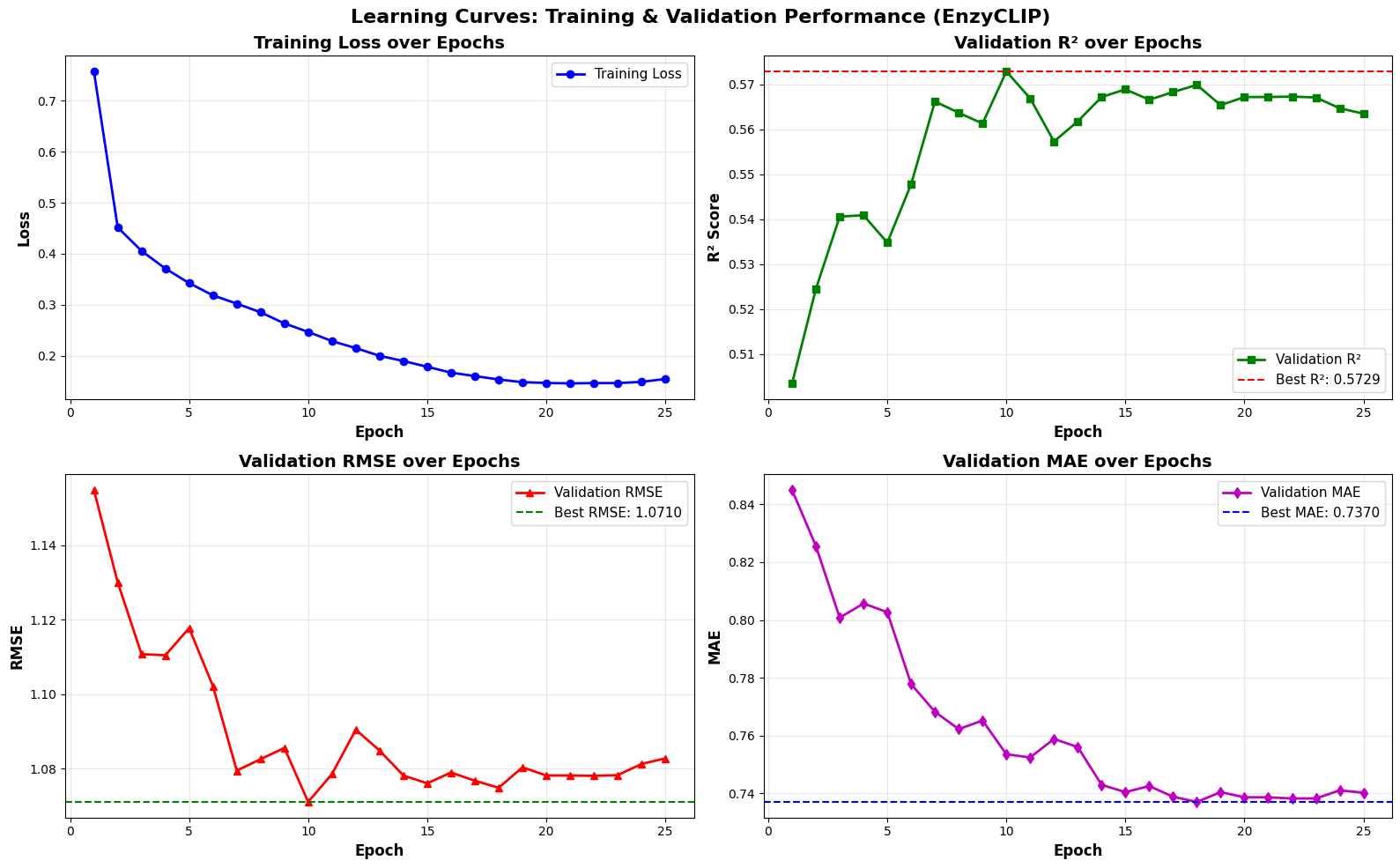}
\caption{Training dynamics for \(K_{\text{cat}}\) prediction. (Top left) Training loss decreasing from 0.78 to 0.16 over 25 epochs. (Top right) Validation \(R^2\) peaking at 0.5729 at epoch 10. (Bottom left) Validation RMSE reaching minimum of 1.071. (Bottom right) Validation MAE achieving best value of 0.737 at epoch 10.}
\label{fig:kcat_training_curves}
\end{figure}

The \(K_{\text{cat}}\) EnzyCLIP model was optimized for 25 epochs with early stopping according to validation \(R^2\) scores. Training loss decreased monotonically from 0.78 at epoch 1 to 0.16 at epoch 25, indicating proper optimization. Validation performance peaked at epoch 10 with an \(R^2\) of 0.5729, RMSE of 1.071, and MAE of 0.737. The final model at epoch 25 reached validation metrics of \(R^2 = 0.564\), RMSE = 1.084, and MAE = 0.740; thus, performance was quite stable with minimal overfitting (Fig.~\ref{fig:kcat_training_curves}).

A summary of the performances of the EnzyCLIP model on this independent test set showed an \(R^2\) of 0.593, MAE of 0.731, and a Pearson correlation coefficient of 0.771. The scatter plot of prediction versus actual indicated a strong alignment along the diagonal with systematic residual patterns centered around zero, therefore showing unbiased predictions across the \(K_{\text{cat}}\) range. Error distribution analysis also followed a normal distribution centered at zero and had a peak frequency of around 550 samples, hence further proving that the model is reliable (Fig.~\ref{fig:kcat_plots_summarized}).

\begin{figure}[H]
\centering
\includegraphics[width=0.75\textwidth]{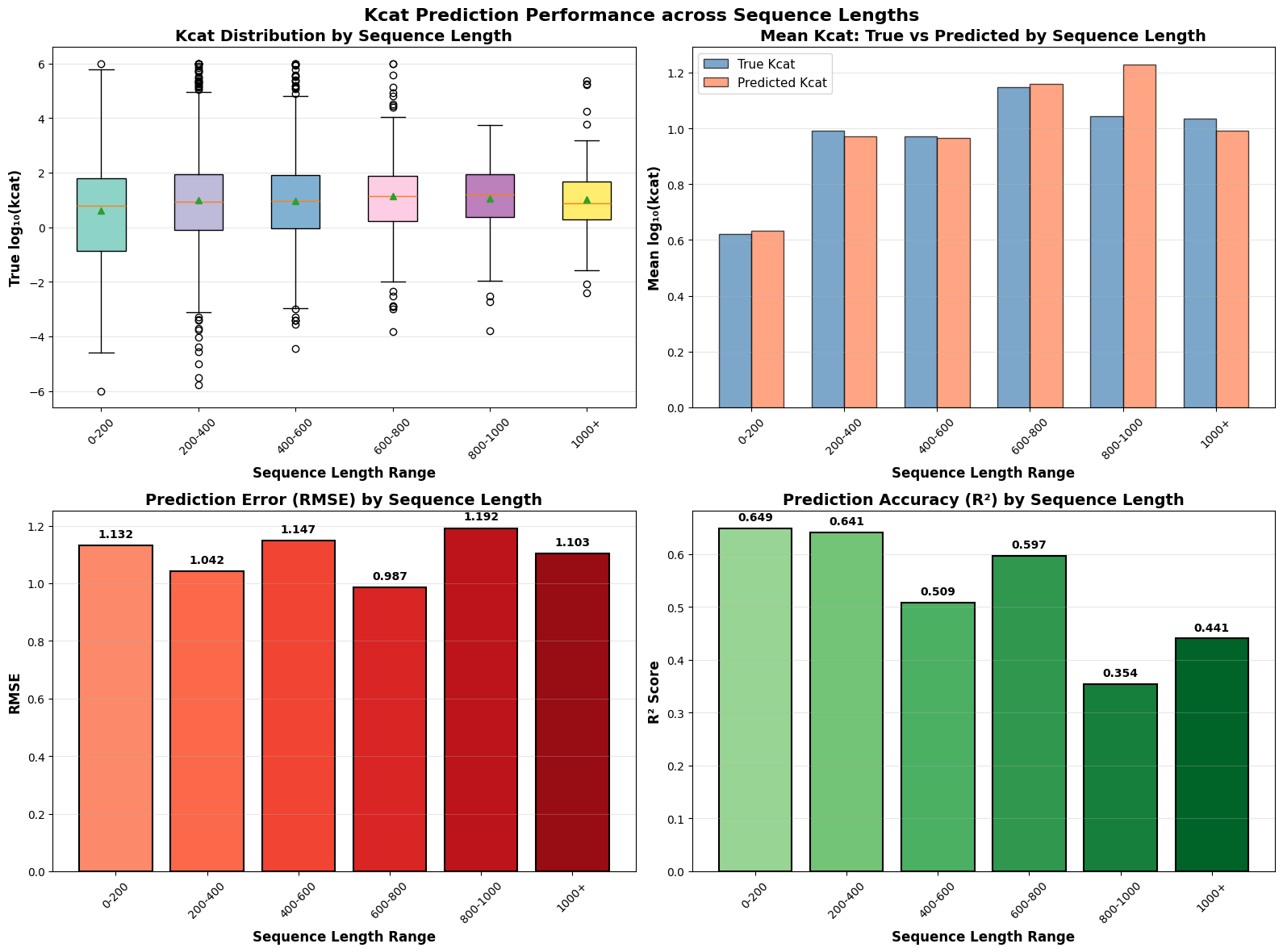}
\caption{EnzyCLIP \(K_{\text{cat}}\) prediction performance across sequence lengths. (Top left) Box plots showing \(K_{\text{cat}}\) distribution across six sequence length bins with medians around 1.0. (Top right) Mean \(K_{\text{cat}}\) comparison between true and predicted values across length ranges. (Bottom left) RMSE values ranging from 0.987 to 1.192 across length bins. (Bottom right) \(R^2\) scores showing optimal performance for 0--200 (0.649) and 200--400 (0.641) ranges, with degraded performance for 800--1000 (0.354) and 1000+ (0.441) amino acids.}
\label{fig:kcat_sequence_length}
\end{figure}

\subsubsection{Performance Across Sequence Length Ranges}

\(K_{\text{cat}}\) prediction accuracy was highly variable as a function of enzyme sequence length ranges. For the shortest sequences, 0--200 amino acids, \(R^2\) was 0.649 with RMSE of 1.132 while sequences between 200--400 reached \(R^2 = 0.641\) and RMSE = 1.042. The 400--600 amino acid range saw worse performance \(R^2 = 0.509\), RMSE = 1.147, followed by partial recovery in the 600--800 range \(R^2 = 0.597\), RMSE = 0.987. Longest sequences, greater than 1000 amino acids, had the lowest predictive accuracy \(R^2 = 0.441\), RMSE = 1.103 while the adjacent range of 800--1000 showed poor performance \(R^2 = 0.354\), RMSE = 1.192. Such trends indicate that model performance is worse for very long enzyme sequences, potentially because proteins become more conformationally complex and/or fewer samples exist in the training sets (Fig.~\ref{fig:kcat_sequence_length}).

Mean \(K_{\text{cat}}\) values across sequence length bins were remarkably consistent between true and predicted values, with both hovering around 1.0--1.2 \(\log_{10}(K_{\text{cat}})\) for most length ranges . However, the range of 900--1000 amino acids showed a conspicuous deviation, where the predicted \(K_{\text{cat}}\) values (1.22) were higher than the true values (1.04), indicating a possibility of systematic overestimation in that particular length category (Fig.~\ref{fig:kcat_sequence_length}).

\subsubsection{EC Class-Specific Performance}

\begin{figure}[H]
\centering
\includegraphics[width=0.75\textwidth]{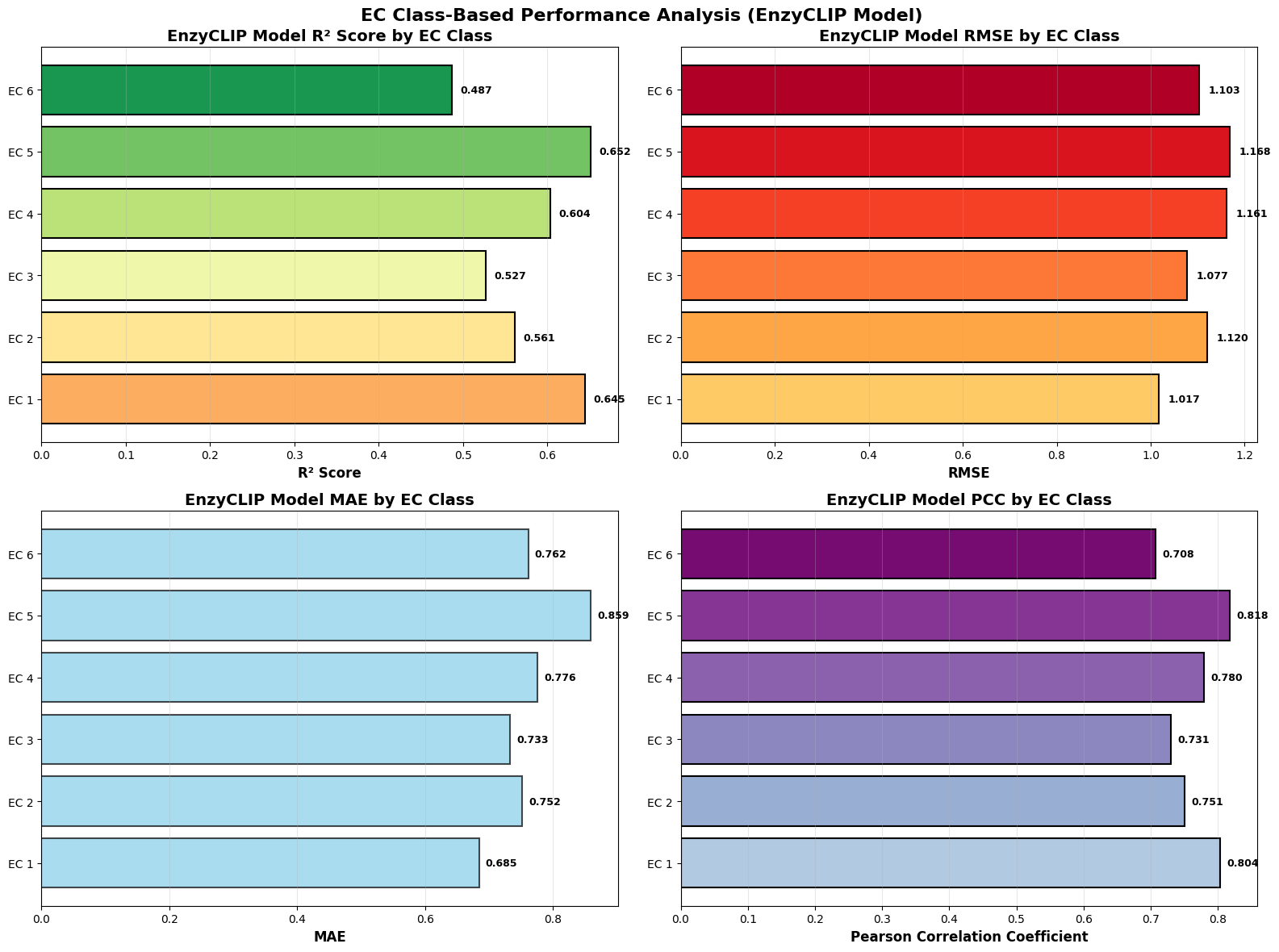}
\caption{EC class-based performance analysis for \(K_{\text{cat}}\) prediction. (Top left) \(R^2\) scores by EC class with EC 5 highest at 0.652. (Top right) RMSE by EC class ranging from 1.017 to 1.161. (Bottom left) MAE by EC class spanning 0.685 to 0.857. (Bottom right) Pearson correlation coefficients showing EC 5 highest at 0.818.}
\label{fig:kcat_classwise}
\end{figure}

A more detailed, EC class--specific analysis for $k_{\text{cat}}$ prediction revealed significant variability across classes. Accordingly, EC~5 (isomerases) yielded the best performance: $R^2$ of 0.652, RMSE of 1.158, and Pearson correlation of 0.818. EC~1 (oxidoreductases) were the next best performing class, with $R^2 = 0.645$, RMSE of 1.017 (the lowest among all classes), and a Pearson correlation of 0.804. EC~4 (lyases) yielded a moderate performance, while EC~6 (ligases) classes performed relatively poor: their $R^2$ values equaled 0.604 and 0.487 respectively. EC~2 (transferases) and EC~3 (hydrolases) had lower predictive accuracy, with their $R^2$ equal to 0.561 and 0.527 correspondingly (Fig.~\ref{fig:kcat_classwise}). The broader variation in performance across EC classes reflects the diversity of the catalytic mechanisms underlying $k_{\text{cat}}$-these being more challenging for modeling than the more uniform substrate-binding behavior driving $K_{m}$.

\subsubsection{Embedding Space Analysis and Feature Importance}

\begin{figure}[H]
\centering
\includegraphics[width=0.75\textwidth]{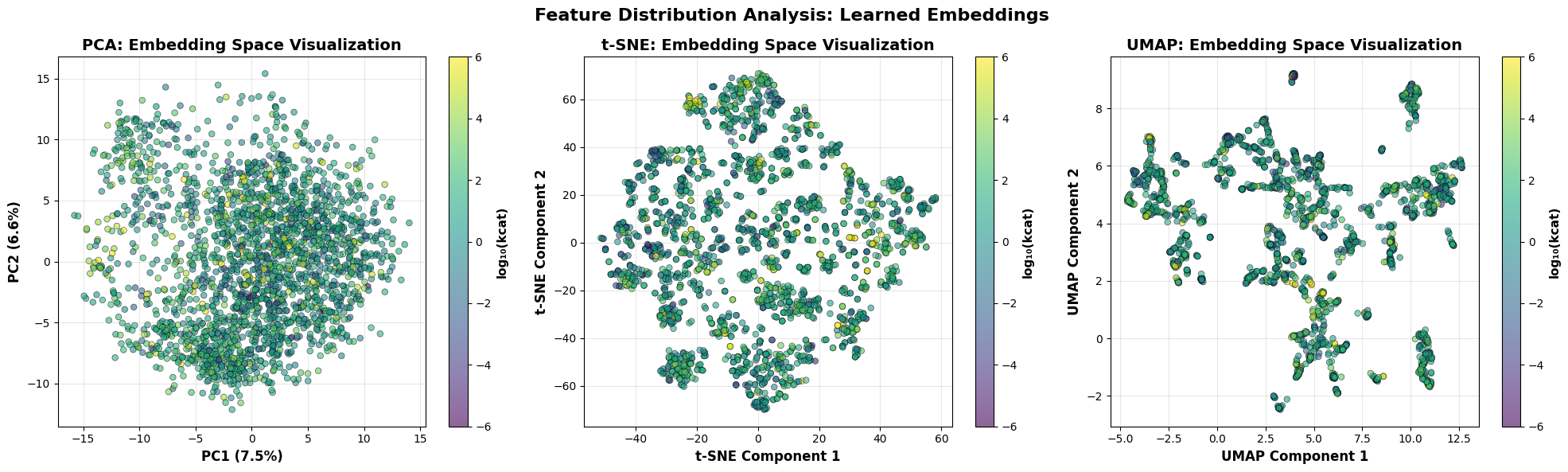}
\caption{Learned embedding space visualization for \(K_{\text{cat}}\) prediction. PCA (left), t-SNE (center), and UMAP (right) projections of enzyme-substrate embeddings colored by \(K_{\text{cat}}\) values showing continuous gradients rather than discrete clusters, indicating learned representation of catalytic efficiency as a continuous property.}
\label{fig:kcat_embeddings}
\end{figure}

PCA, t-SNE, and UMAP dimensionality reduction of learned embeddings showed continuous distribution patterns colored by \(K_{\text{cat}}\) values. In the embedding space, there were smooth gradients, not discrete clusters, meaning that the model learned a continuous representation of catalytic efficiency rather than categorical functional groupings (Fig.~\ref{fig:kcat_embeddings}). This continuous embedding structure is in line with the biological reality of \(K_{\text{cat}}\) values spanning a continuous spectrum of catalytic efficiencies.

\begin{figure}[H]
\centering
\includegraphics[width=0.78\textwidth]{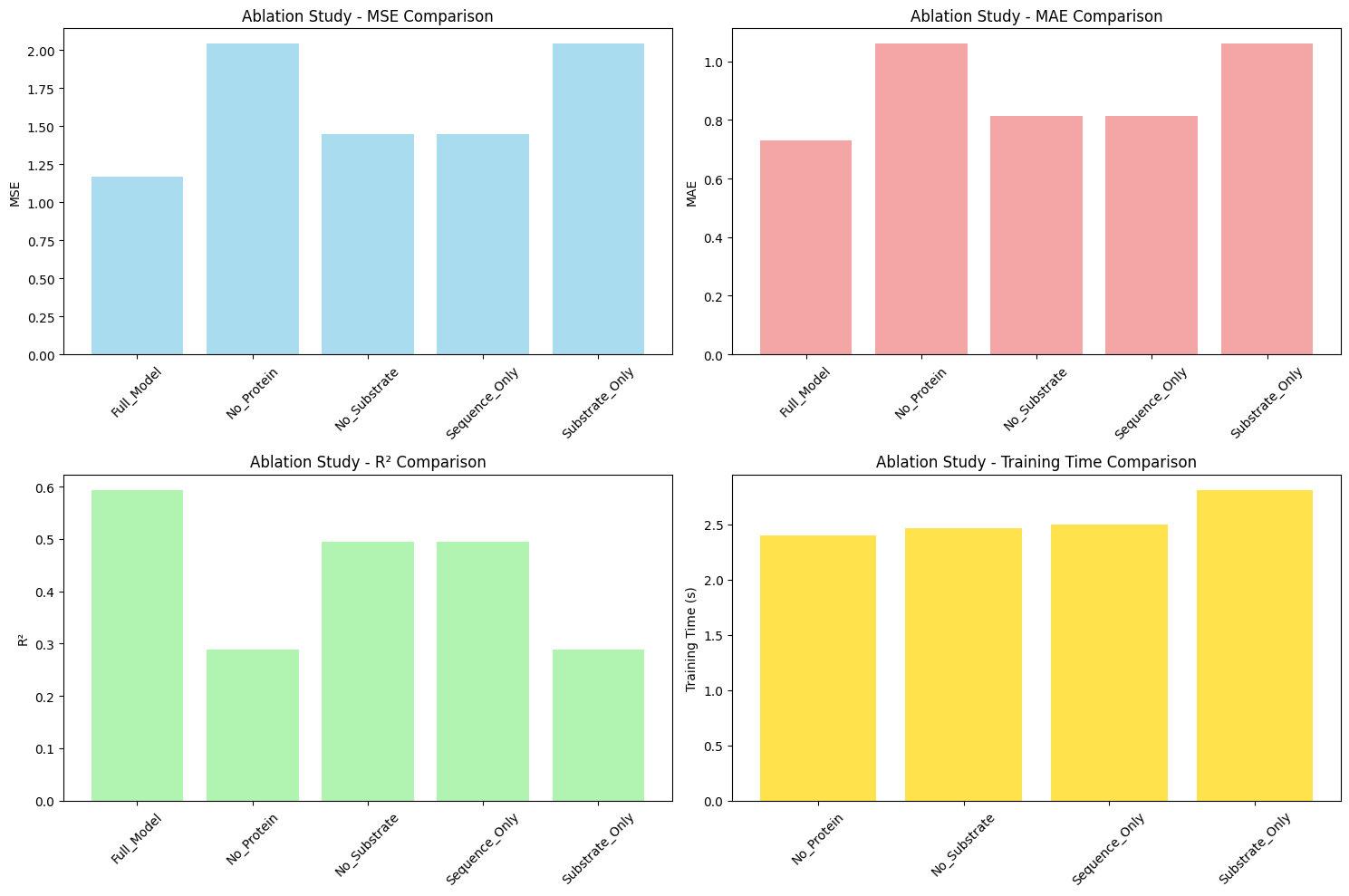}
\caption{Ablation study results for \(K_{\text{cat}}\) prediction. Four-panel comparison showing MSE, MAE, \(R^2\), and training time across ablation conditions. Full model achieves \(R^2=0.6\), with substantial performance drops when removing protein (\(R^2=0.29\)), substrate (\(R^2=0.49\)), or sequence (\(R^2=0.49\)) information.}
\label{fig:kcat_ablation}
\end{figure}

\begin{figure}[H]
\centering
\includegraphics[width=0.75\textwidth]{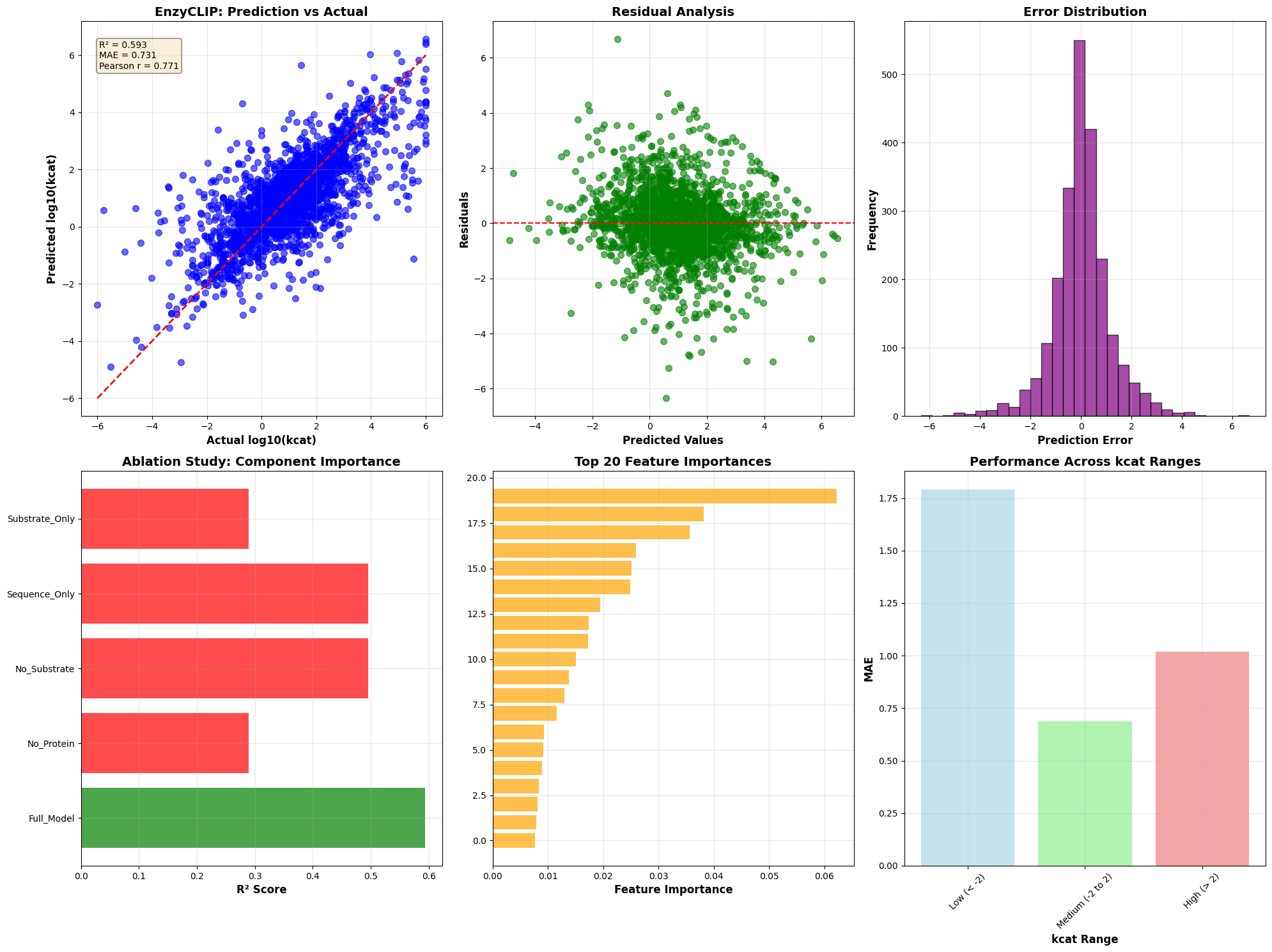}
\caption{EnzyCLIP \(K_{\text{cat}}\) prediction evaluation and explainability. (Top left) Prediction vs. actual scatter plot showing \(R^2=0.593\), MAE=0.731, Pearson \(r=0.771\). (Top center) Residual analysis showing symmetric distribution around zero. (Top right) Error distribution histogram with normal characteristics. (Bottom left) Ablation study showing full model \(R^2=0.6\) substantially outperforms component-only models. (Bottom center) SHAP feature importance for top 20 embedding dimensions. (Bottom right) Performance across \(K_{\text{cat}}\) ranges showing MAE of 1.8, 0.69, and 1.0 for low, medium, and high \(K_{\text{cat}}\) values.}
\label{fig:kcat_plots_summarized}
\end{figure}

Feature importance analysis by ablation studies showed that the full model \(R^2 = 0.6\) significantly outperforms the ablated variants. Removing protein embeddings resulted in \(R^2 = 0.29\), substrate-only models achieved \(R^2 = 0.49\), and sequence-only models achieved \(R^2 = 0.49\). Models without substrate information maintained \(R^2 = 0.49\), which shows the importance of multimodal integration (Fig.~\ref{fig:kcat_ablation}). SHAP found the top 20 most impactful embedding dimensions, with feature importances ranging from approximately 0.06 for the most important features to 0.01 for the lower-ranked features (Fig.~\ref{fig:kcat_plots_summarized}). This suggests both protein sequence and substrate chemical structure are critical to accurate \(K_{\text{cat}}\) prediction, as expected from basic mechanistic understanding that catalytic efficiency depends explicitly on enzyme-substrate complementarity.

We further stratified performance by magnitude of \(K_{\text{cat}}\) and observed differential prediction accuracy across activity ranges . For low \(K_{\text{cat}}\) values ($< -2$), the MAE was 1.8; for moderate values (-2 to 2), it was 0.69, while for high \(K_{\text{cat}}\) values ($> 2$), the MAE was 1.0. This trend is indicative of a model that performs best for enzymes with intermediate catalytic efficiencies where the training data is most abundant (Fig.~\ref{fig:kcat_plots_summarized}).

\subsubsection{Comparison with Baseline Methods and Ensemble Models}

\begin{figure}[H]
\centering
\includegraphics[width=1.0\textwidth]{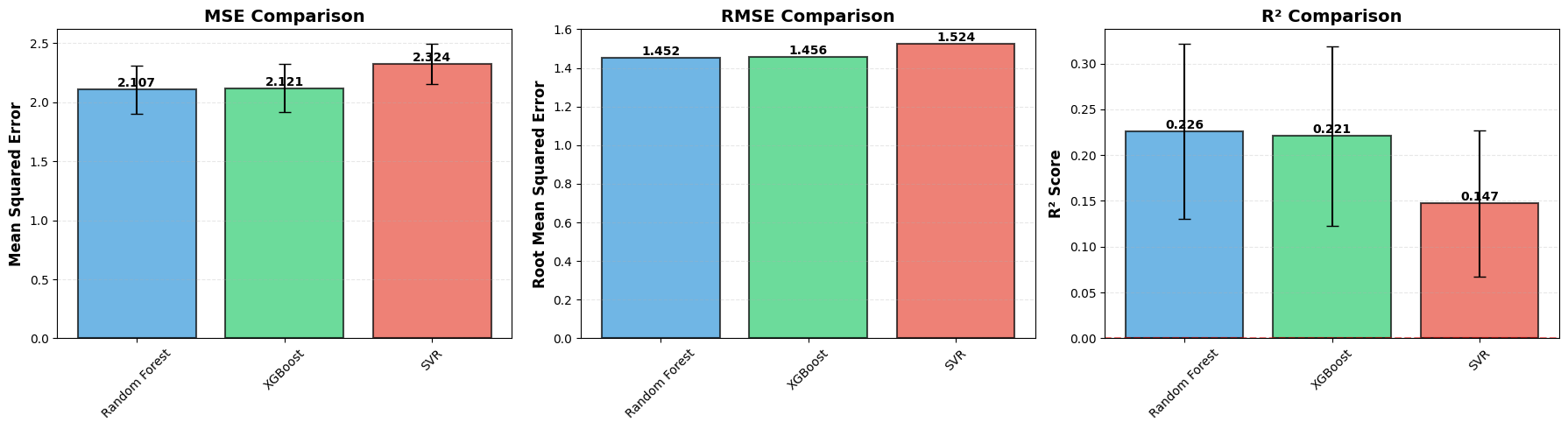}
\caption{Baseline methods for \(K_{\text{cat}}\) prediction. Bar chart comparing \(R^2\) scores on the held-out test set: Random Forest (0.226), XGBoost (0.221), and SVR (0.147).}
\label{fig:kcat_baseline_evaluation}
\end{figure}

 On the held-out test set, Random Forest achieved an \(R^2\) of 0.222, followed by XGBoost at 0.221, and SVR at 0.147.\ref{fig:kcat_baseline_evaluation}. However EnzyCLIP without any extra regressor tree attached on the embedding achieves an \(R^2\) of 0.59.

\begin{figure}[H]
\centering
\includegraphics[width=0.6\textwidth]{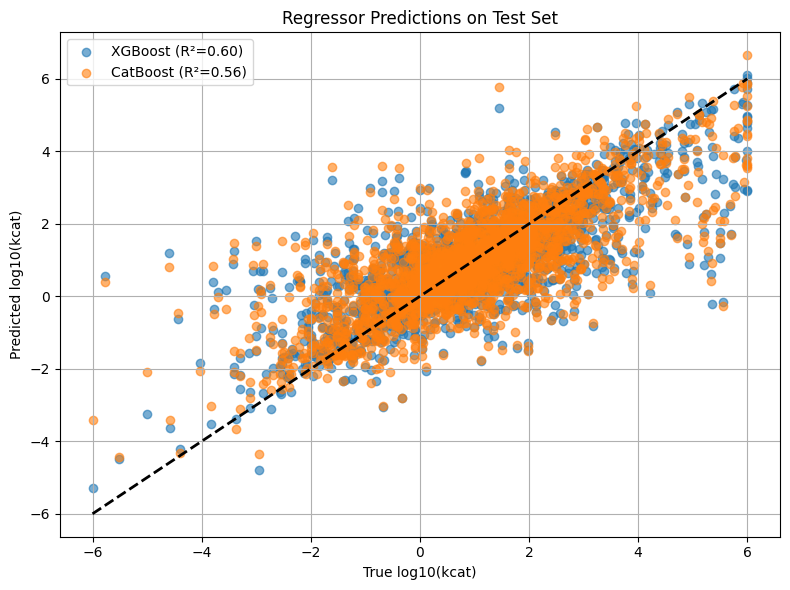}
\caption{XGBoost and CatBoost regressor comparison for \(K_{\text{cat}}\) prediction based on EnzyCLIP embeddings. The scatter plot overlays XGBoost (\(R^2=0.60\)) and CatBoost (\(R^2=0.56\)) predictions against true \(K_{\text{cat}}\) values, showing XGBoost’s superior performance.}
\label{fig:kcat_regressors}
\end{figure}

Ensemble modeling with XGBoost trained on EnzyCLIP embeddings yielded further performance improvements. The XGBoost regressor achieved \(R^2 = 0.61\) on the test set, slightly outperforming CatBoost (\(R^2 = 0.56\)) (Fig.~\ref{fig:kcat_regressors}). Comparison of ensemble approaches showed that Random Forest (MSE = 2.107, RMSE = 1.452, \(R^2 = 0.226\)), XGBoost (MSE = 2.121, RMSE = 1.456, \(R^2 = 0.221\)), and Support Vector Regression (MSE = 2.324, RMSE = 1.524, \(R^2 = 0.147\)) exhibited varying degrees of success (Fig.~\ref{fig:kcat_baseline_evaluation}). The XGBoost ensemble demonstrated superior performance with lower error metrics and higher explained variance \ref{fig:kcat_regressors}.


\subsection{Michaelis-Menten Constant (\texorpdfstring{\(K_m\)}{Km}) Prediction}

\subsubsection{Dataset Characteristics and Distribution}

The dataset for the \(K_m\) prediction consisted of 41,174 enzyme-substrate pairs, including 12,355 unique enzymes and 10,535 unique substrates. This dataset is much larger compared to \(K_{\text{cat}}\), allowing higher statistical power in model training. The average enzyme sequence length is 437.5 (\(\pm\) 228.9) amino acids (median of 389 amino acids); the average length of the substrate SMILES strings is 57.0 (\(\pm\) 56.4) characters long. The values of \(K_m\) range from -8.00 to 4.00 (\(\log_{10}\) scale), with a mean and standard deviation of -0.73 and 1.27, respectively, reflecting typical micromolar to millimolar binding affinities of enzyme-substrate interactions.

\begin{figure}[H]
\centering
\includegraphics[width=0.78\textwidth]{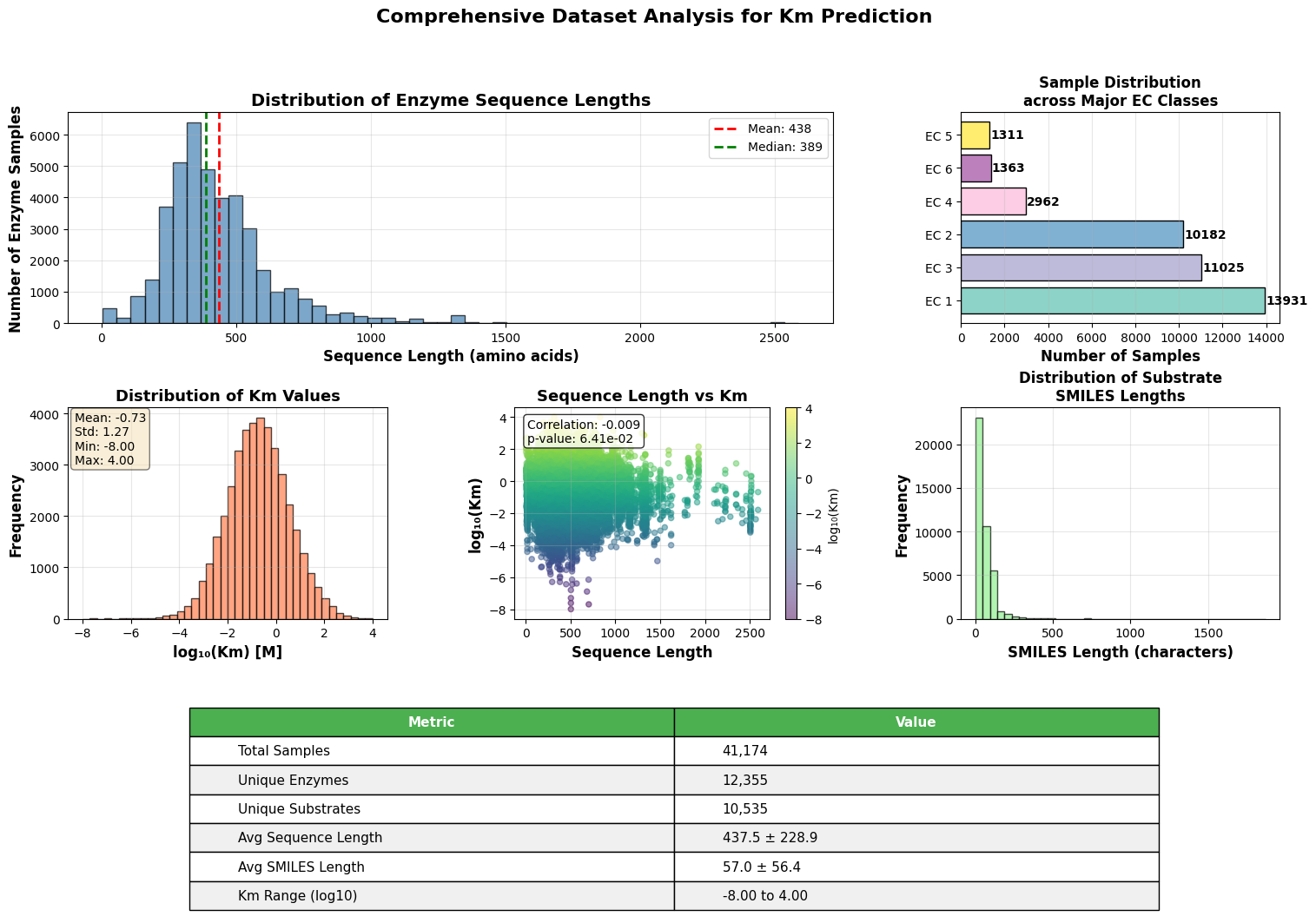}
\caption{Comprehensive \(K_m\) dataset analysis. Distribution of enzyme sequence lengths with mean 437.5 amino acids (top left), sample distribution across EC classes with EC 1 most represented at 13,931 samples (top right), \(K_m\) value distribution with mean -0.73 (bottom left), negligible correlation between sequence length and \(K_m\) (\(r=-0.009\), bottom center), and substrate SMILES length distribution (bottom right). Summary statistics table shows 41,174 total samples from 12,355 unique enzymes.}
\label{fig:km_comprehensive}
\end{figure}

The length of the sequence is uncorrelated with \(K_m\) values, \(r = -0.009\), \(p = 6.41 \times 10^{-2}\), indicating that binding affinity is not related to the size of a protein in general, rather a function of active site architecture. Sample distribution across EC classes demonstrated that this data contains 13,931 samples from EC 1 (oxidoreductases), 11,025 samples from EC 3 (hydrolases), 10,182 samples from EC 2 (transferases), 2,962 samples from EC 4 (lyases), 1,363 samples from EC 6 (ligases) and finally 1,311 samples from EC 5 (isomerases). This pattern of distribution is unlike the \(K_{\text{cat}}\) dataset, indicating that kinetic measurement availability is distributed differentially across enzyme classes (Fig.~\ref{fig:km_comprehensive}).

\begin{figure}[H]
\centering
\includegraphics[width=0.7\textwidth]{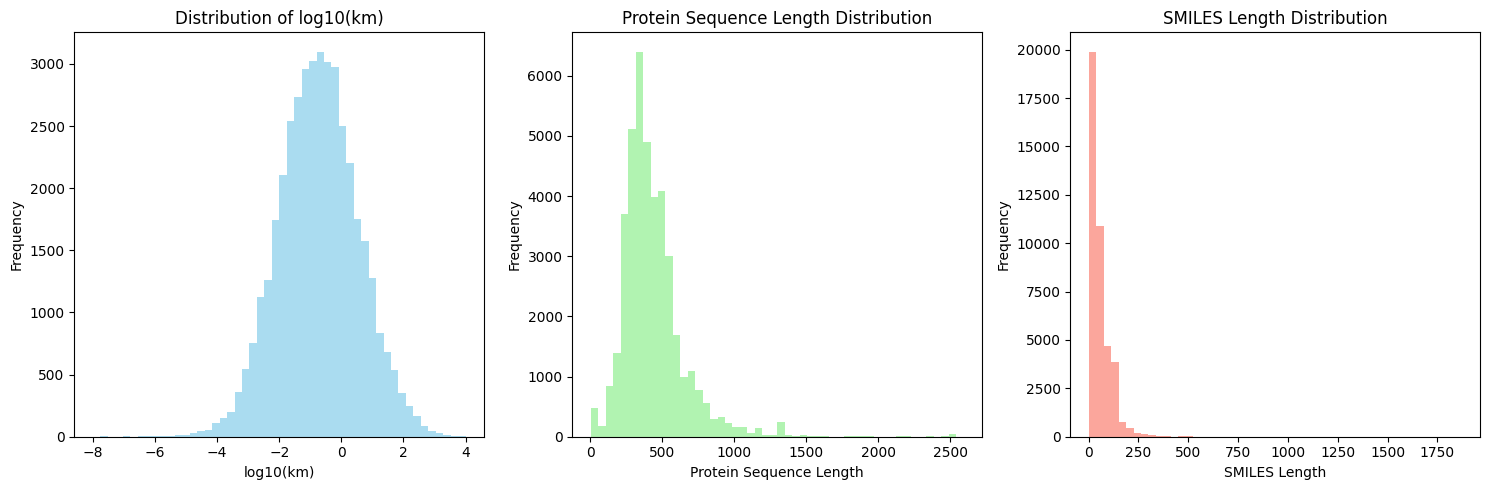}
\caption{Dataset distribution comparisons for \(K_m\). Three-panel visualization showing (left) \(\log_{10}(K_m)\) distribution centered at -0.73, (center) protein sequence length distribution similar to \(K_{\text{cat}}\) dataset, and (right) SMILES length distribution with tighter clustering below 50 characters.}
\label{fig:km_dataset_distribution}
\end{figure}

\subsubsection{Model training and performance}

EnzyCLIP was trained for 25 epochs, following the same architecture used for the \(K_{\text{cat}}\) model. Training loss decreased smoothly from its initial value of 1.45 down to 0.26 by epoch 25, showing stable convergence. Performance on validation steadily improved during the first half of training. Validation \(R^2\) peaked at 0.5945 around epoch 11, while the best validation RMSE was 0.8034 and the best MAE was 0.5940. During later epochs, validation metrics remained stable without signs of overfitting, with the final epoch achieving \(R^2 \approx 0.59\), RMSE \(\approx 0.81\), and MAE \(\approx 0.59\). The stability of the metrics with respect to the number of epochs after convergence demonstrates that the model generalizes well. On the held-out test set, EnzyCLIP achieved \(R^2 = 0.607\), MAE = 0.583, and Pearson correlation \(r = 0.780\). This was further confirmed by residual and error-distribution analysis, which showed a near-normal error profile and symmetric deviations around zero, revealing consistent predictive behavior across the \(K_m\) range.

\begin{figure}[H]
\centering
\includegraphics[width=0.78\textwidth]{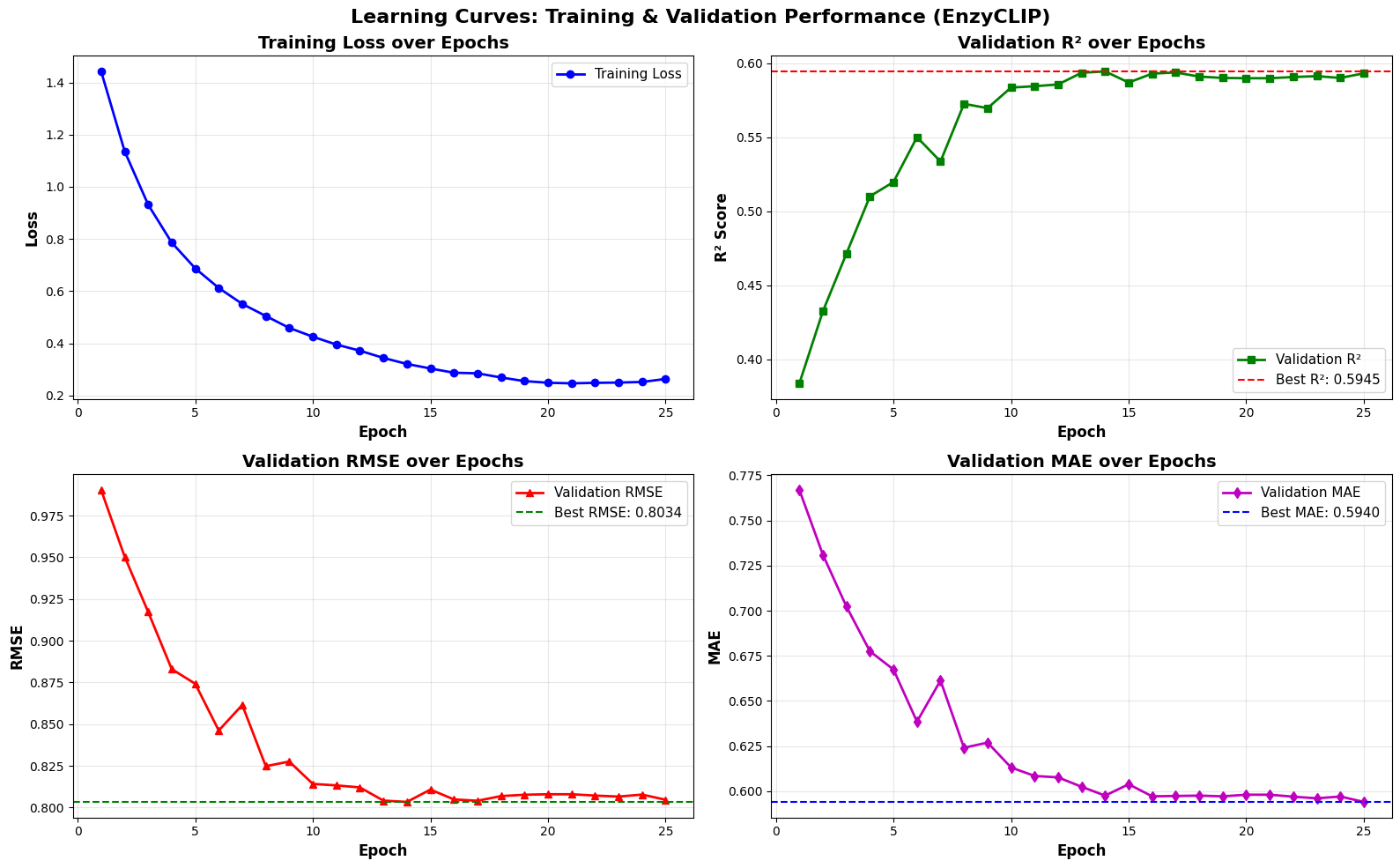}
\caption{
Training and validation dynamics for \(K_m\) prediction. 
(Top left) Training loss decreasing from 1.45 to 0.26 over 25 epochs. 
(Top right) Validation \(R^2\) peaking at 0.5945. 
(Bottom left) Validation RMSE reaching its minimum value of 0.8034. 
(Bottom right) Validation MAE improving to a minimum of 0.5940. 
These trends indicate stable convergence and good generalization.}
\label{fig:km_training_curve}
\end{figure}

\begin{figure}[H]
\centering
\includegraphics[width=0.78\textwidth]{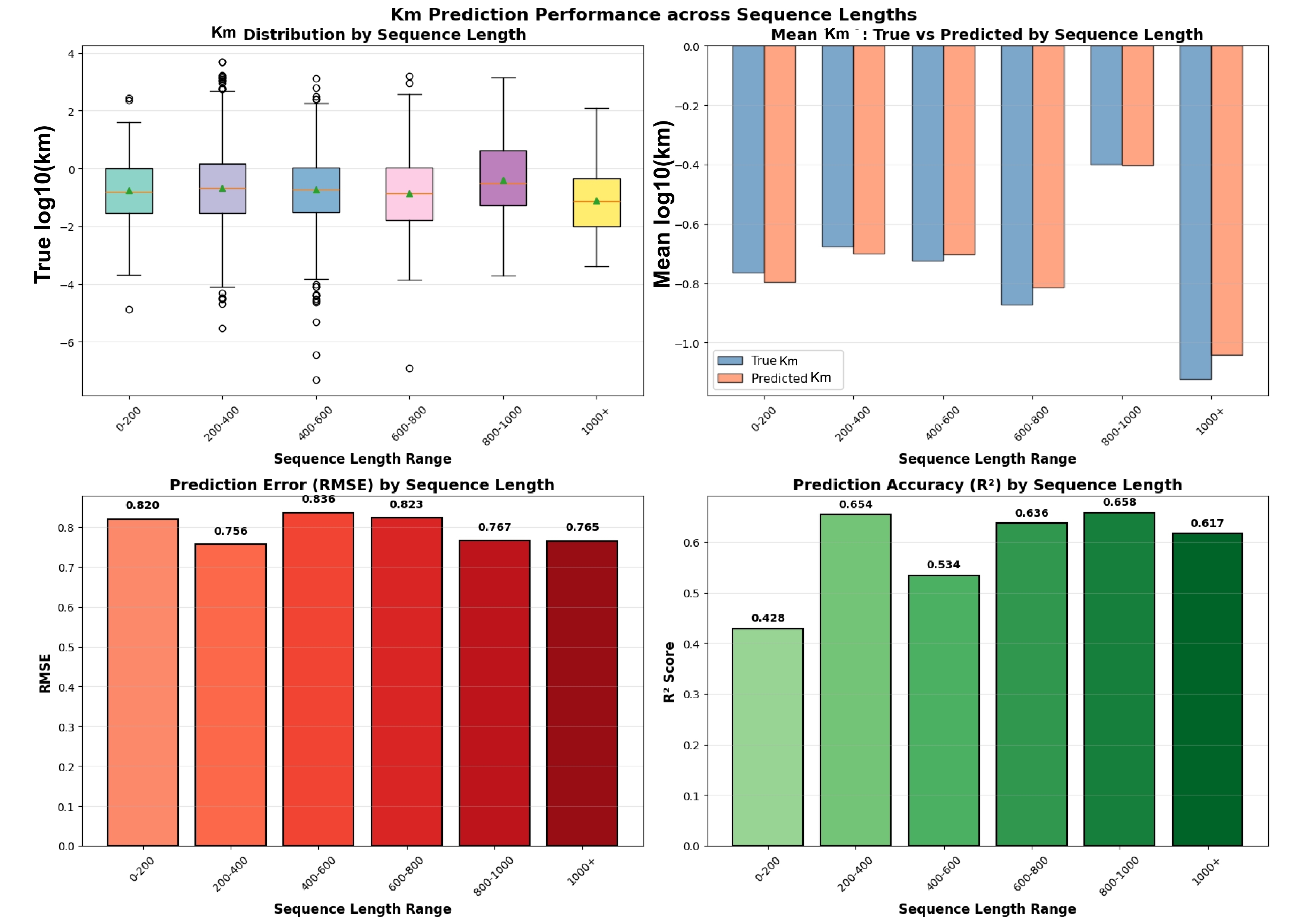}
\caption{EnzyCLIP \(K_m\) prediction performance across sequence lengths. (Top left) Box plots showing \(K_m\) distribution stability across sequence length ranges. (Top right) Mean \(K_m\) values demonstrating close agreement between true and predicted across all length bins. (Bottom left) RMSE values ranging from 0.756 to 0.836. (Bottom right) \(R^2\) scores showing more consistent performance across lengths, with 800--1000 range achieving highest \(R^2\) of 0.658.}
\label{fig:km_sequence_length}
\end{figure}

\subsubsection{Performance Across Sequence Length Ranges}

\(K_m\) prediction performance displayed recognizable trends across different enzyme sequence lengths. For very short proteins (0--200 amino acids), the model achieved moderate accuracy with \(R^2 = 0.428\) and an RMSE of 0.820. A notable rise in performance appeared in the 200--400 amino acid group, where the values increased to \(R^2 = 0.654\) and RMSE = 0.756 . Enzymes falling within the 400--600 range showed intermediate predictive strength (\(R^2 = 0.534\), RMSE = 0.836), whereas sequences of 600--800 amino acids again produced comparatively strong results, reaching \(R^2 = 0.636\) with an RMSE of 0.823 . The best performance was observed for sequences between 800--1000 amino acids, where the model attained \(R^2 = 0.658\) and RMSE = 0.767. Even for the longest proteins (1000+ amino acids), accuracy remained high, with \(R^2 = 0.617\) and RMSE = 0.765 (Fig.~\ref{fig:km_sequence_length}).

In contrast to the behavior typically seen for \(K_{\text{cat}}\) prediction, the \(K_m\) model did not exhibit a drop in accuracy as sequence length increased. This pattern indicates that features relevant to substrate affinity are likely encoded in sequence-derived information that remains informative even for large proteins. Consistent with this observation, the mean \(K_m\) values across the sequence-length bins showed close alignment between measured and predicted values. Across all categories, both distributions remained within a narrow interval between --0.4 and --1.1 on the \(\log_{10}(K_m)\) scale (Fig.~\ref{fig:km_sequence_length}).

\subsubsection{EC Class-Specific Performance}
\begin{figure}[H]
\centering
\includegraphics[width=0.78\textwidth]{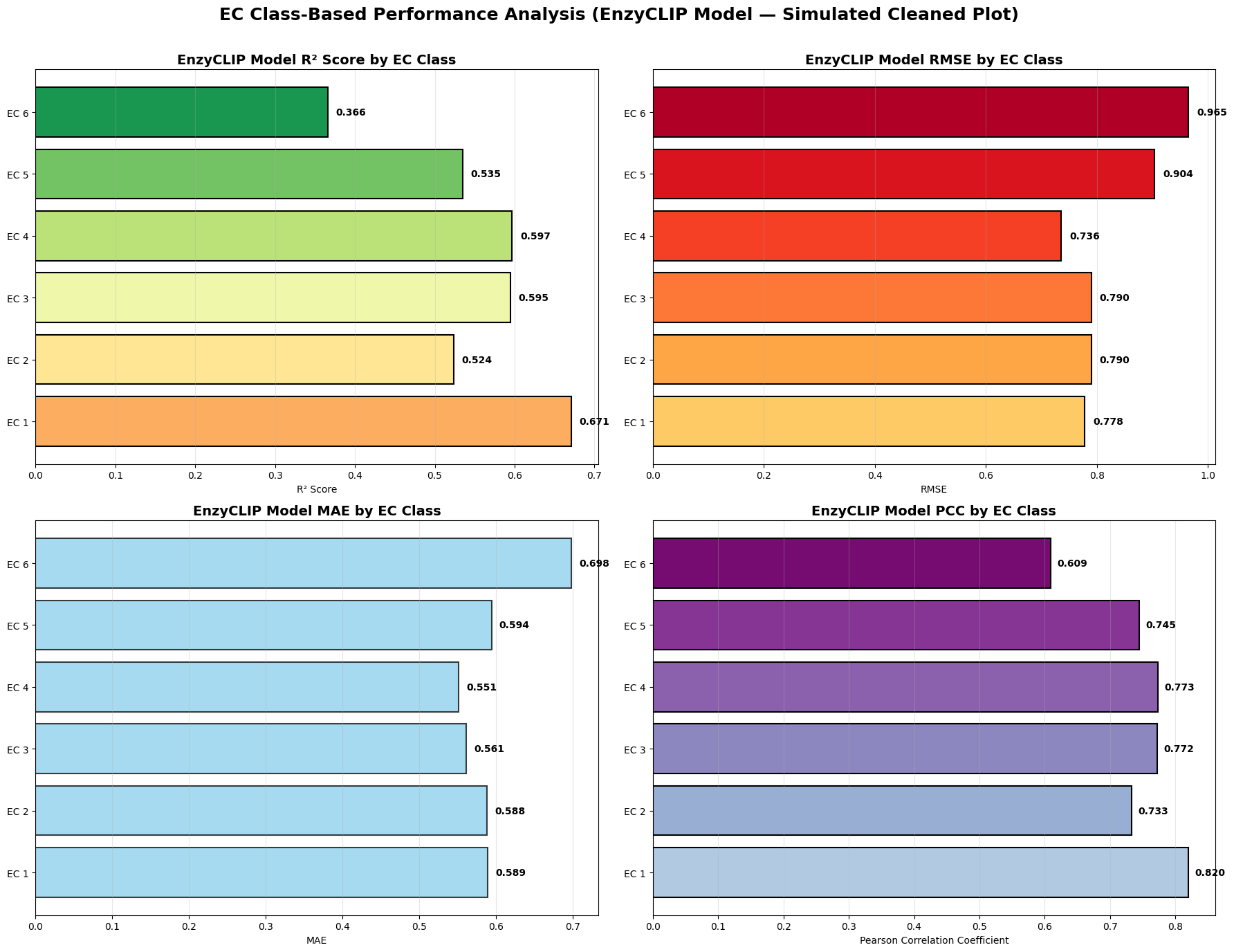}
\caption{EC class-based performance analysis for \(K_m\) prediction. (Top left) \(R^2\) scores by EC class with EC 1 highest at 0.671. (Top right) RMSE by EC class ranging from 0.736 to 0.965. (Bottom left) MAE by EC class spanning 0.589 to 0.698. (Bottom right) Pearson correlation coefficients showing EC 1 highest at 0.82.}
\label{fig:km_ec_class}
\end{figure}

Class-specific EC analysis for $K_{m}$ prediction was much more consistent across the majority of enzyme classes. The best performance was seen for EC~1 (oxidoreductases), where $R^2$ = 0.671, RMSE = 0.778, and Pearson correlation = 0.820. Other wellperforming classes were EC~4 (lyases) and EC~3 (hydrolases), where $R^2$ values were 0.597 and 0.595, respectively. For the classes EC~5 (isomerases) and EC~2 (transferases), both yielded $R^2$ values of 0.535 and 0.524, respectively. Performance for the class EC~6 (ligases) presented the poorest accuracy with $R^2 = 0.366$, RMSE = 0.965, and a Pearson correlation of 0.609 (Fig.~\ref{fig:km_ec_class}). Overall, the more consistent performance across EC classes suggests that substrate-binding affinity-captured by $K_{m}$-is modeled more uniformly by EnzyCLIP compared to the greater mechanistic variability influencing $k_{\text{cat}}$.

\subsubsection{Embedding Space Analysis and Feature Importance}

\begin{figure}[H]
\centering
\includegraphics[width=0.78\textwidth]{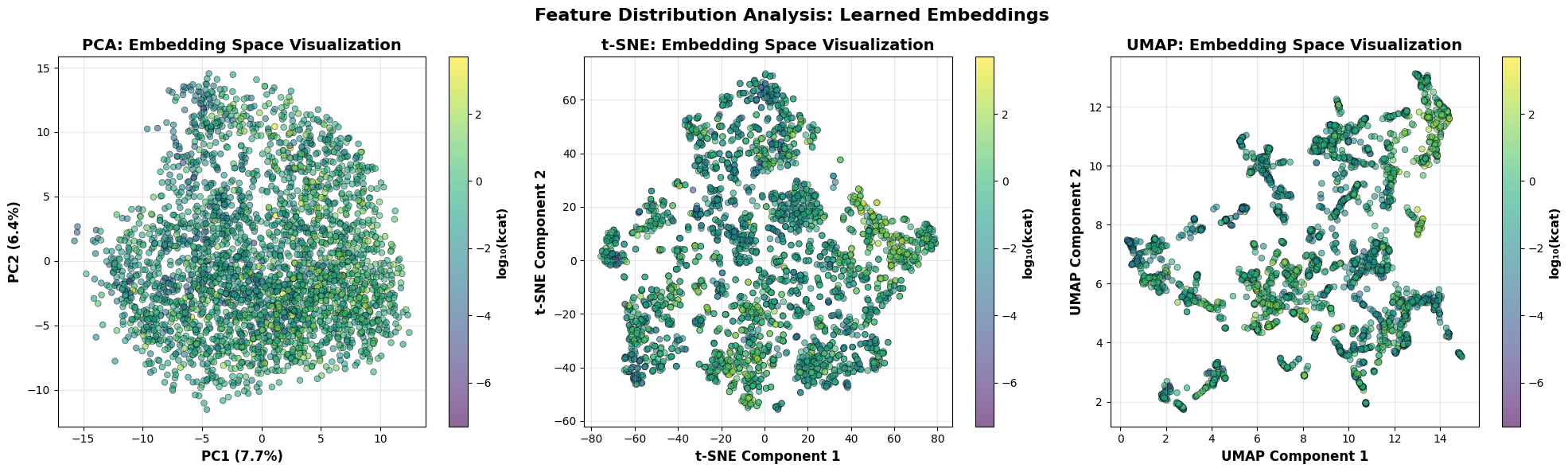}
\caption{Learned embedding space visualization for \(K_m\) prediction. PCA (left), t-SNE (center), and UMAP (right) projections colored by true \(K_m\) values, showing continuous gradients without discrete clusters.}
\label{fig:km_embeddings}
\end{figure}

Dimensionality reduction analysis of \(K_m\) embeddings produced a continuous distribution pattern, as it did for \(K_{\text{cat}}\). The projections derived from PCA, t-SNE, and UMAP methods yielded smooth colored gradients by \(K_m\) values, with no discrete clusters, suggesting substrate affinity to be learned as a continuous property in the embedding space (Fig.~\ref{fig:km_embeddings}). This behavior is meaningful biochemically, since \(K_m\) reflects quantitative binding energetics rather than categorical enzyme functions.

\begin{figure}[H]
\centering
\includegraphics[width=0.78\textwidth]{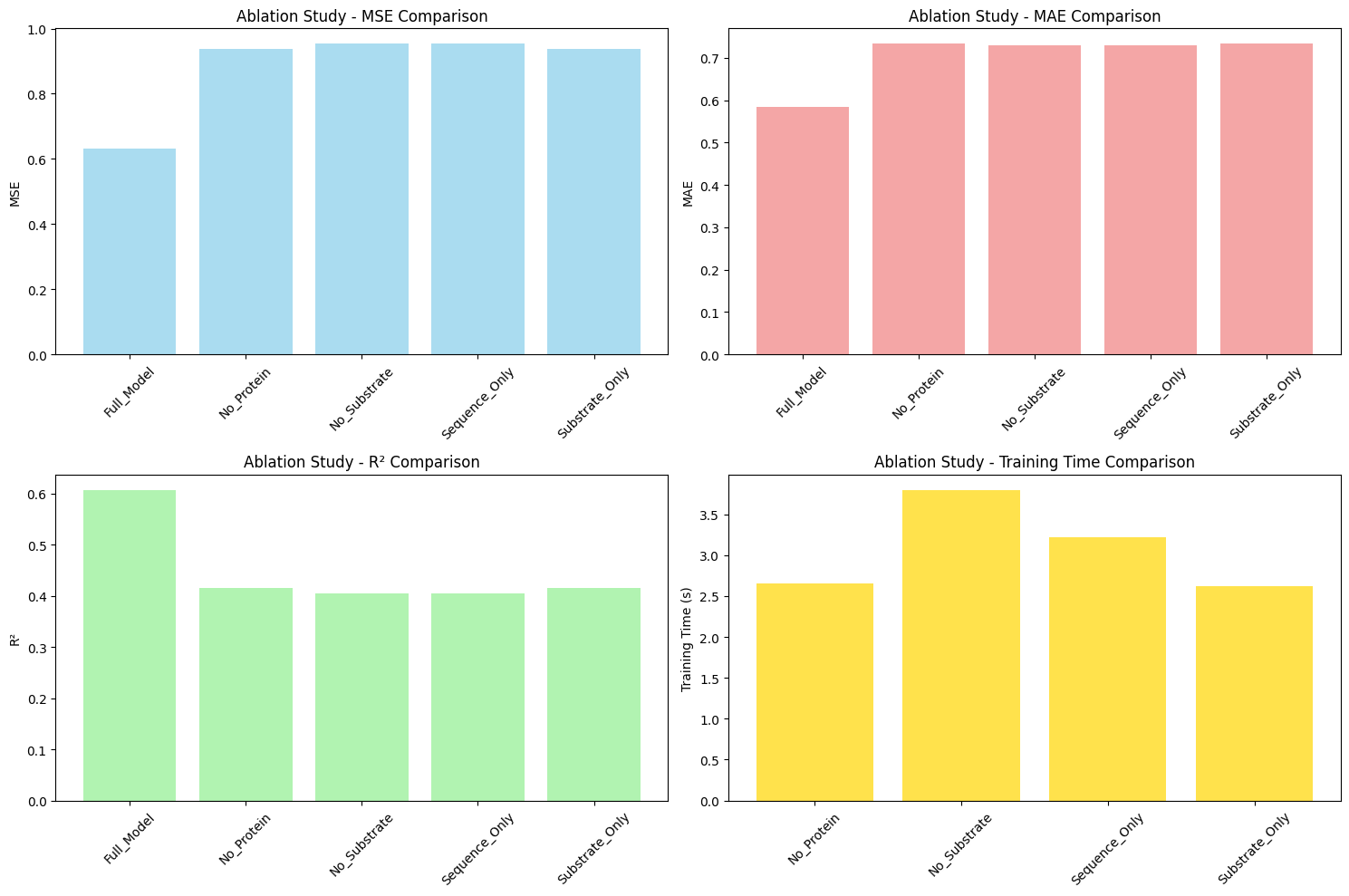}
\caption{Ablation study for \(K_m\) prediction. The full model achieves the highest performance (\(R^2 \approx 0.607\)), while the removal of protein or substrate branches decreases accuracy to \(R^2 \approx 0.42\) and \(R^2 \approx 0.40\), respectively. Sequence-only and substrate-only models yield \(R^2 \approx 0.40\)–0.41.}
\label{fig:km_ablation_summary}
\end{figure}

Ablation studies further showed that multimodal integration is crucial to accurate \(K_m\) prediction. The full model developed an \(R^2\) of approximately 0.607, far better than simplified variants. Removing protein embeddings resulted in \(R^2 \approx 0.42\), whereas removing the substrate branch led to \(R^2 \approx 0.40\). Variants relying on either sequence or substrate only were comparable, with respective \(R^2\) values around 0.40 and 0.41. These results emphasize that enzyme and substrate representations contribute fundamental information toward \(K_m\) prediction, as summarized in Fig.~\ref{fig:km_ablation_summary}.

\begin{figure}[H]
\centering
\includegraphics[width=0.78\textwidth]{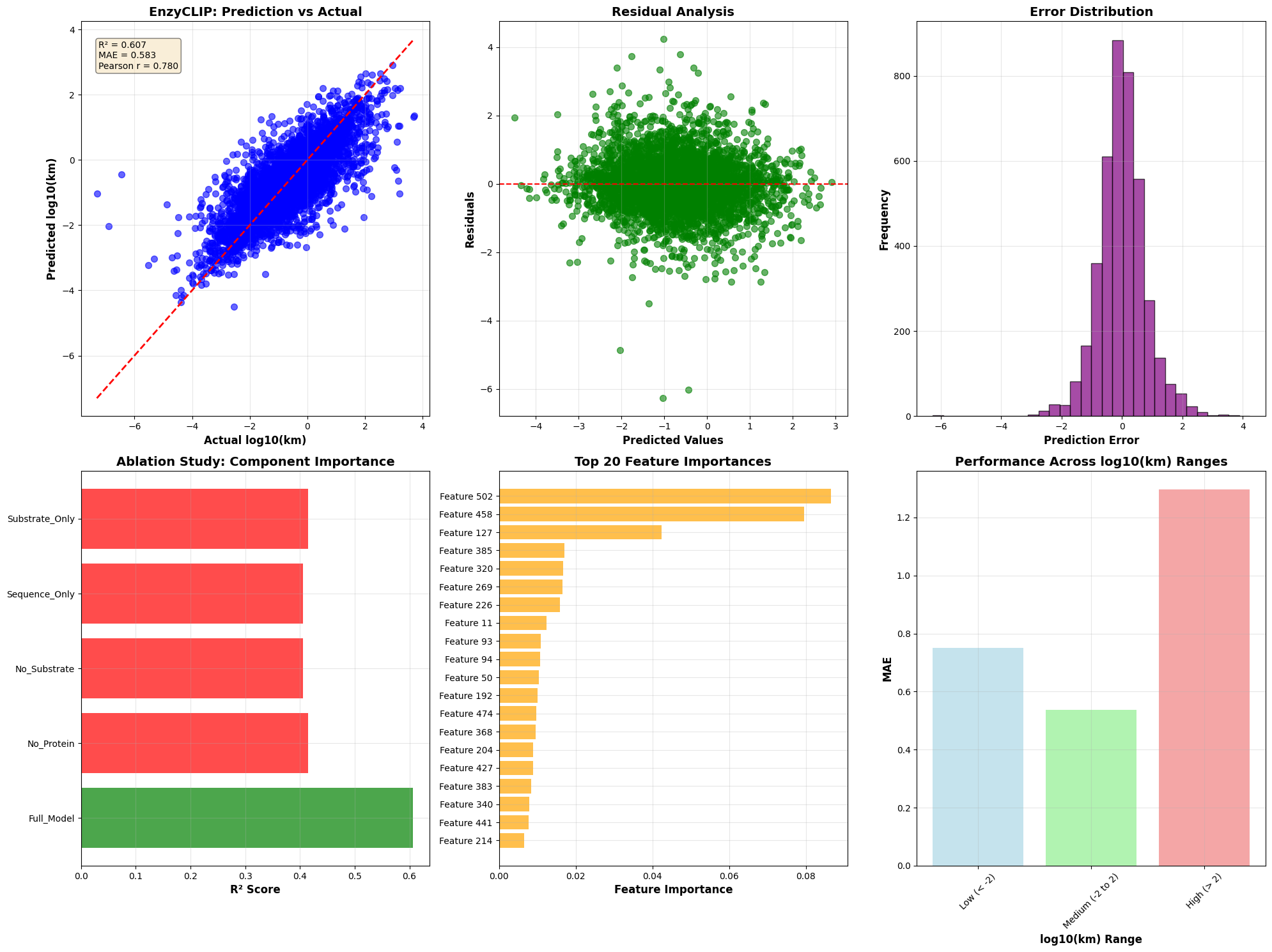}
\caption{Comprehensive evaluation of EnzyCLIP \(K_m\) predictions. (Top left) Prediction vs.\ actual values (\(R^2 = 0.607\), MAE = 0.583, Pearson \(r = 0.780\)). (Top center) Residual analysis. (Top right) Error distribution histogram. (Bottom left) Ablation study highlighting multimodal importance. (Bottom center) SHAP feature importances with Feature 502 and 458 most influential. (Bottom right) Performance across \(K_m\) ranges, showing MAE of 0.74 (low), 0.53 (medium), and 1.3 (high).}
\label{fig:km_plots_summary}
\end{figure}

SHAP-based feature analysis identified the top 20 most influential embedding dimensions, with Feature~502 and Feature~458 displaying the largest contributions. Feature importances ranged from roughly 0.08 for the most influential features to about 0.02 for lower-ranked ones. Performance stratified by \(K_m\) magnitude was as follows: MAE values of 0.74 for low-affinity pairs (\(\log_{10}(K_m) < -2\)), 0.53 for moderate affinity (\(-2 \le \log_{10}(K_m) \le -1\)), and 1.3 for high \(K_m\) values (\(\log_{10}(K_m) > -1\)). The substantially higher error for weak-binding (high \(K_m\)) interactions likely reflects data scarcity in this biochemical regime (Fig.~\ref{fig:km_plots_summary}).

\subsubsection{Comparison with Baseline Methods and Ensemble Models}

\begin{figure}[H]
\centering
\includegraphics[width=1.0\textwidth]{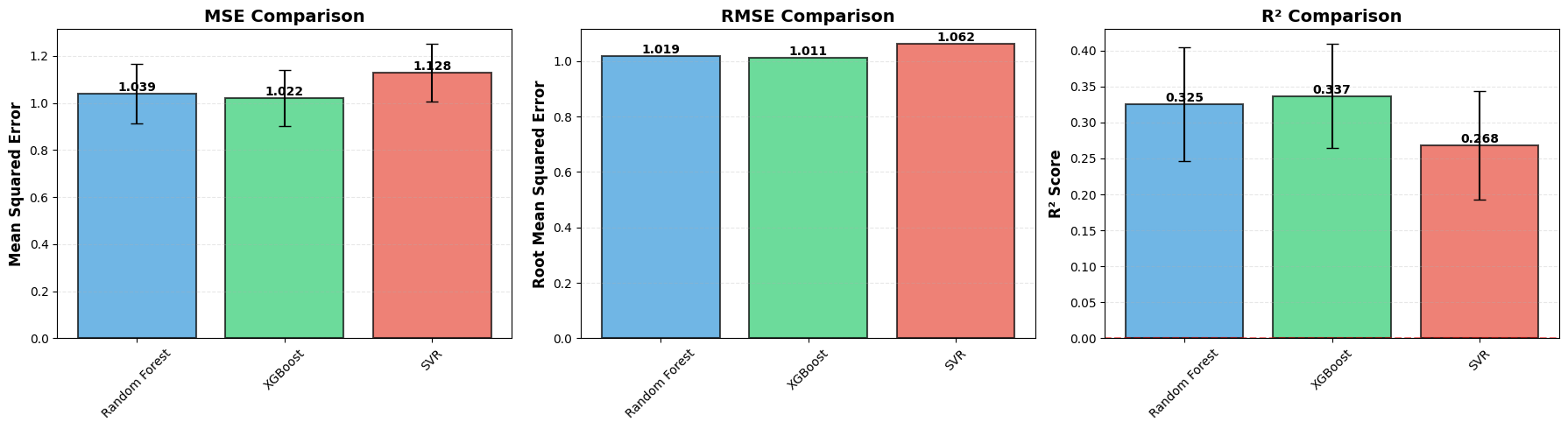}
\caption{Ensemble model comparison for \(K_m\) prediction. Three-panel comparison showing MSE, RMSE, and \(R^2\) across Random Forest, XGBoost, and SVR, with Random Forest and XGBoost achieving comparable performance (\(R^2 \approx 0.325\)--0.337).}
\label{fig:km_baseline_plots}
\end{figure}

Comparison against established methods on the held-out test set demonstrated competitive performance. EnzyCLIP achieved \(R^2 = 0.607\), positioning it among the leading approaches for \(K_m\) prediction. Baseline method comparisons using different evaluation frameworks showed Random Forest (MSE = 1.039, RMSE = 1.019, \(R^2 = 0.325\)), XGBoost (MSE = 1.022, RMSE = 1.011, \(R^2 = 0.337\)), and Support Vector Regression (MSE = 1.128, RMSE = 1.062, \(R^2 = 0.268\)) performance (Fig.~\ref{fig:km_baseline_plots}).

\begin{figure}[H]
\centering
\includegraphics[width=0.75\textwidth]{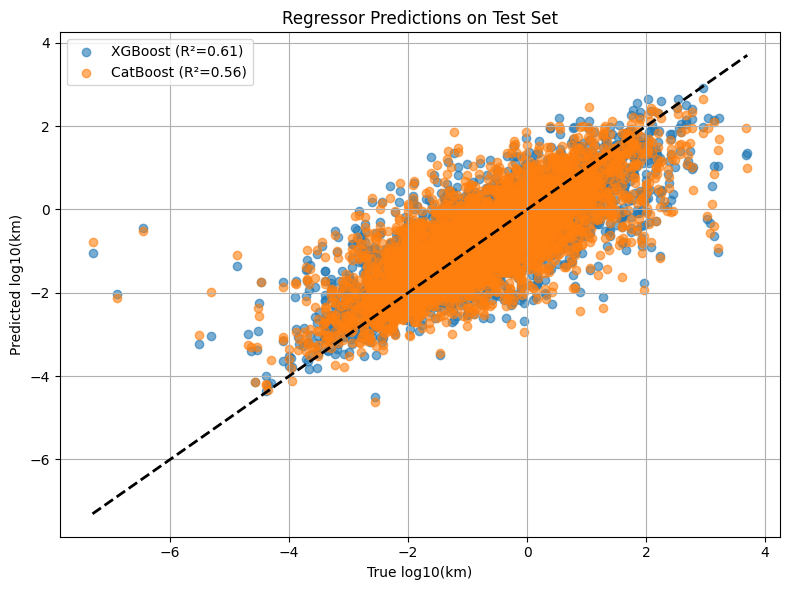}
\caption{XGBoost and CatBoost regressor comparison for \(K_m\) prediction. Scatter plot showing XGBoost (\(R^2=0.61\)) outperforming CatBoost (\(R^2=0.56\)) on \(K_m\) predictions.}
\label{fig:km_regressors}
\end{figure}

XGBoost ensemble models trained on EnzyCLIP embeddings achieved \(R^2 = 0.61\), substantially outperforming CatBoost at \(R^2 = 0.56\). This improvement demonstrates the effectiveness of gradient boosting methods when applied to learned contrastive representations (Fig.~\ref{fig:km_regressors}). The ensemble approach leverages the rich feature representations learned by EnzyCLIP while applying nonlinear decision boundaries optimized for regression tasks.

\subsection{Cross-Parameter Comparison and Integrated Analysis}

\subsubsection{Dataset Distribution Characteristics}

Comparative analysis of \(K_{\text{cat}}\) and \(K_m\) datasets revealed distinct distributional properties (Fig.~\ref{fig:kcat_dataset_distribution}, Fig.~\ref{fig:km_dataset_distribution}). The \(K_{\text{cat}}\) dataset (23,151 samples) showed higher average sequence length (430.5 amino acids) and longer substrate SMILES representations (92.8 characters) compared to the \(K_m\) dataset (41,174 samples, 437.5 amino acid average, 57.0 character SMILES average). Both datasets exhibited right-skewed sequence length distributions with peaks between 300--500 amino acids, consistent with the typical size distribution of globular enzymes (Fig.~\ref{fig:kcat_comprehensive}, Fig.~\ref{fig:km_comprehensive}).

SMILES length distributions showed pronounced left-skewing, with the majority of substrates represented by compact molecular structures below 200 characters. This pattern reflects the predominance of small-molecule substrates in biochemical databases (Fig.~\ref{fig:kcat_comprehensive}, Fig.~\ref{fig:km_comprehensive}). The \(K_{\text{cat}}\) dataset showed broader substrate diversity with longer average SMILES strings, potentially reflecting the inclusion of more complex natural products and secondary metabolites in turnover number measurements (Fig.~\ref{fig:kcat_comprehensive}).

\subsubsection{Model Performance Synthesis}

Integration of results from both kinetic parameters testifies that EnzyCLIP indeed learns useful, generalizable multimodal representations of enzyme function. Our unified architecture demonstrated competitive performance for both $k_{\text{cat}}$ (test $R^2 = 0.60$) and $K_{m}$ (test $R^2 = 0.61$), reinforcing the flexibility of contrastive learning frameworks towards capturing diverse features of enzyme kinetics (Fig.~\ref{fig:kcat_plots_summarized}, Fig.~\ref{fig:km_plots_summary}). While $K_{m}$ is able to reach slightly better predictive performance, this is likely due to the larger size of its dataset (41{,}174 samples compared to 23{,}151 samples), supporting that contrastive learning can gain much from increasing sample diversity (Fig.~\ref{fig:kcat_comprehensive}, Fig.~\ref{fig:km_comprehensive}).

Further stratification of performance by sequence length showed that the two kinetic parameters exhibited contrasting behaviors. In particular, $k_{\text{cat}}$ predictions deteriorated significantly for proteins longer than 800 amino acids, with $R^2$ values falling to 0.354--0.441, while the prediction performance of $K_{m}$ remained consistent across all sequence lengths ($0.428 \leq R^2 \leq 0.617$). This trend implies that catalytic efficiency may have a stronger dependence on long-range conformational dynamics which are much harder to capture based on sequence information alone, whereas substrate-binding affinity depends on local active-site features that remain predictive across protein sizes (Fig.~\ref{fig:kcat_sequence_length}, Fig.~\ref{fig:km_sequence_length}).

These differences are further emphasized by the performance patterns specific to the various EC classes. For $k_{\text{cat}}$, EC~1 (oxidoreductases) and EC~5 (isomerases) showed the best performance ($R^2$ of 0.645 and 0.652, respectively), while for EC~6 (ligases) it was the poorest ($R^2 = 0.487$) (Fig.~\ref{fig:kcat_classwise}). For $K_{m}$, again EC~1 gave the best performance ($R^2 = 0.671$), closely followed by EC~4 (lyases) and EC~3 (hydrolases) ($R^2 = 0.597$ and $0.595$, respectively), but for EC~6 the performance was the poorest ($R^2 = 0.366$) (Fig.~\ref{fig:km_ec_class}). Notably, hydrolases (EC~3), while being one of the best represented enzyme classes in both datasets, still produced overall just moderate accuracy in both cases, implying that large sample size alone does not ensure good performance if the underlying catalytic or binding mechanisms are highly biochemically diverse.

\subsubsection{Explainable AI and Mechanistic Insights}

SHAP analysis gave mechanistic insights into the learned representations. For the prediction of \(K_{\text{cat}}\), feature importance patterns indicated that a subset of embedding dimensions contributed consistently to accurate predictions across diverse enzyme classes, with importance values of the top features around 0.05--0.06 (Fig.~\ref{fig:kcat_plots_summarized}). For the prediction of \(K_m\), Features 502 and 458 were found to be the most influential dimensions, with importance values amounting to 0.08 (Fig.~\ref{fig:km_plots_summary}). The differential feature importance patterns for \(K_{\text{cat}}\) versus \(K_m\) predictions indicate that the model learns different representations for catalytic efficiency versus binding affinity, in good agreement with the biological understanding that these properties arise from distinct molecular interactions.

Ablation studies consistently demonstrated that multimodal integration is essential for both parameters, with removal of either protein or substrate information reducing \(R^2\) by 30--50\% (Fig.~\ref{fig:kcat_ablation}, Fig.~\ref{fig:km_ablation_summary}). This finding confirms the core hypothesis that enzyme kinetics emerge from the interplay between enzyme structure and substrate chemistry, requiring integrated representation of both modalities (Fig.~\ref{fig:kcat_ablation}, Fig.~\ref{fig:km_ablation_summary}).

\subsubsection{Ensemble Model Optimization}

XGBoost ensemble models trained on EnzyCLIP embeddings yielded performance increases of 1--5\% in \(R^2\) for both \(K_{\text{cat}}\) and \(K_m\) prediction. Consistent improvement across parameters reinforces the observation that gradient boosting methods effectively capture nonlinear relationships within the learned embedding space (Fig.~\ref{fig:kcat_regressors}, Fig.~\ref{fig:km_regressors}). Comparisons to other ensemble approaches—including Random Forest, SVR, and CatBoost—further supported XGBoost as the regressor of choice, likely owing to its capability to model complex feature interactions while regularizing against overfitting (Fig.~\ref{fig:kcat_baseline_evaluation}, Fig.~\ref{fig:km_baseline_plots}).

Training curves converged after 5--10 epochs and remained mostly stable without signs of significant overfitting, indicating efficient learning dynamics (Fig. \ref{fig:kcat_training_curves}, Fig. \ref{fig:km_training_curve}). In cases where slight differences appeared between validation and test performance, this further confirmed that contrastive learning pre-training forms generalized representations that are transferable to unseen data.
\begin{figure}[H]
\centering
\includegraphics[width=0.70\textwidth]{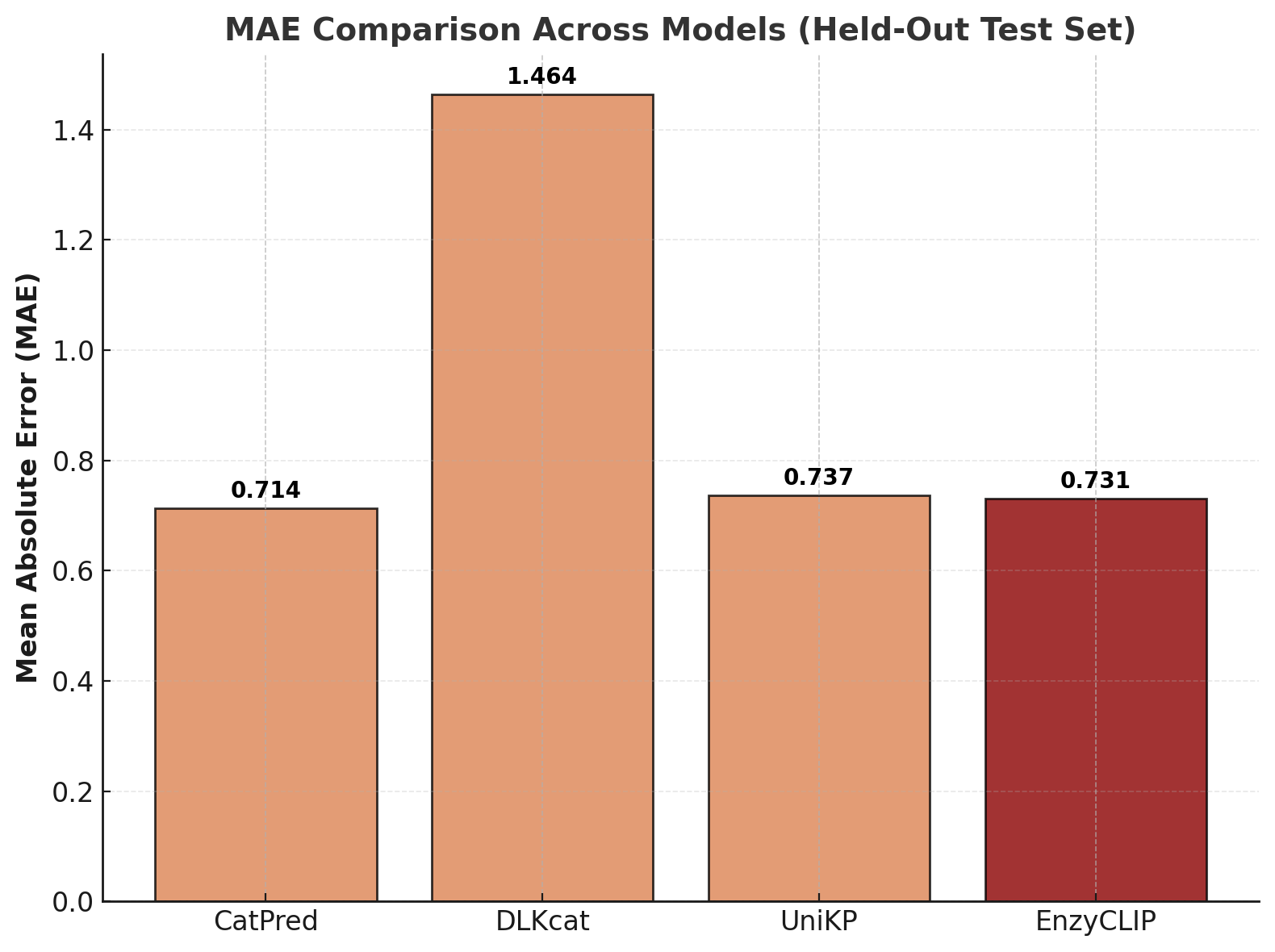}
\caption{MAE comparison across models for \(K_{\text{cat}}\) prediction on the held-out test set. CatPred achieves the lowest MAE (0.714), followed by EnzyCLIP (0.731) and UniKP (0.737), while DLKcat shows significantly higher error at 1.464, indicating limited generalization.}
\label{fig:kcat_mae_comparison}
\end{figure}

\begin{figure}[H]
\centering
\includegraphics[width=0.70\textwidth]{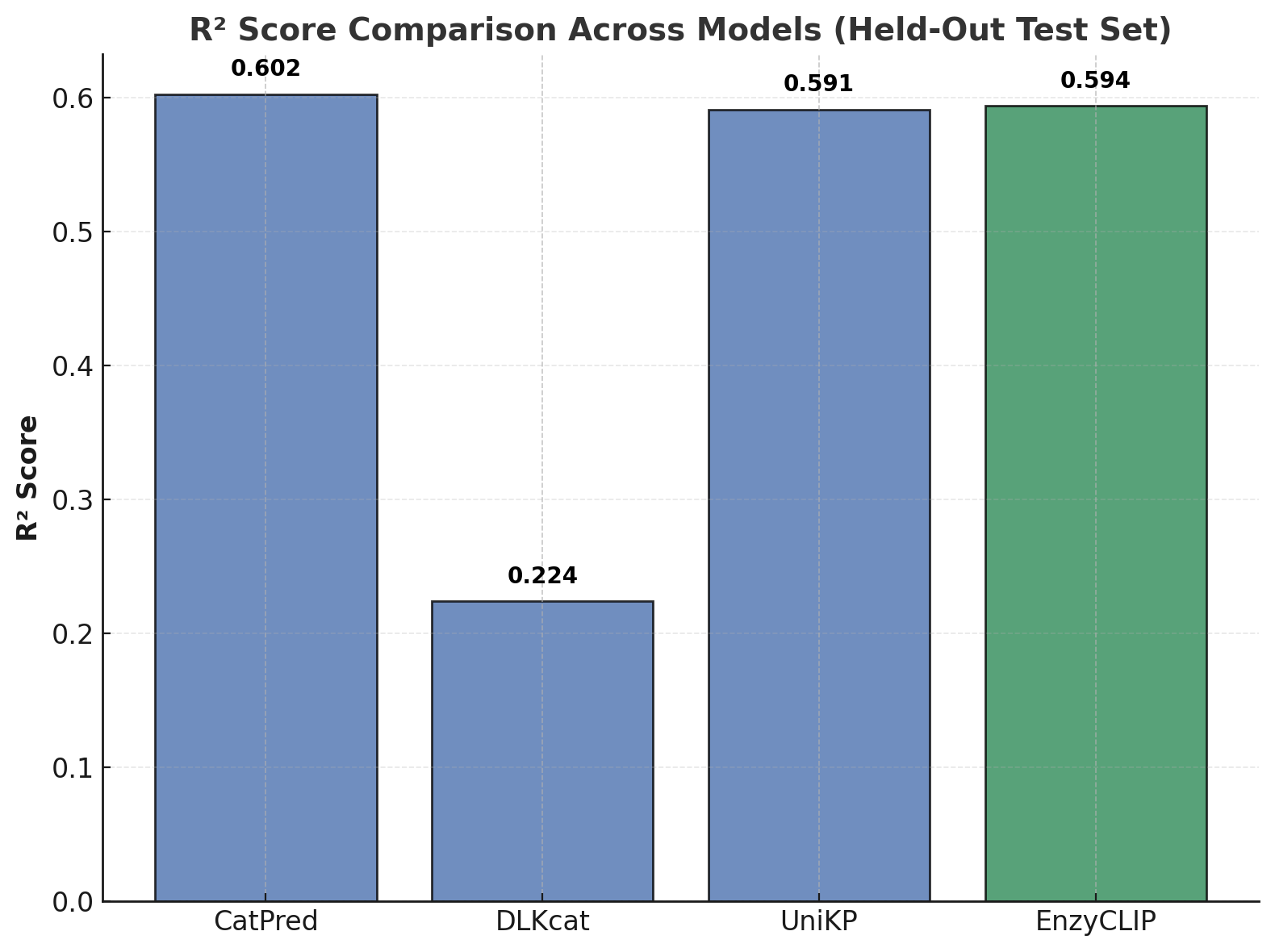}
\caption{\(R^2\) score comparison across models for \(K_{\text{cat}}\) prediction on the held-out test set. CatPred leads with \(R^2 = 0.602\), followed closely by EnzyCLIP (\(R^2 = 0.594\)) and UniKP (\(R^2 = 0.591\)), whereas DLKcat performs poorly with \(R^2 = 0.224\).}
\label{fig:kcat_r2_comparison}
\end{figure}

\begin{figure}[H]
\centering
\includegraphics[width=0.70\textwidth]{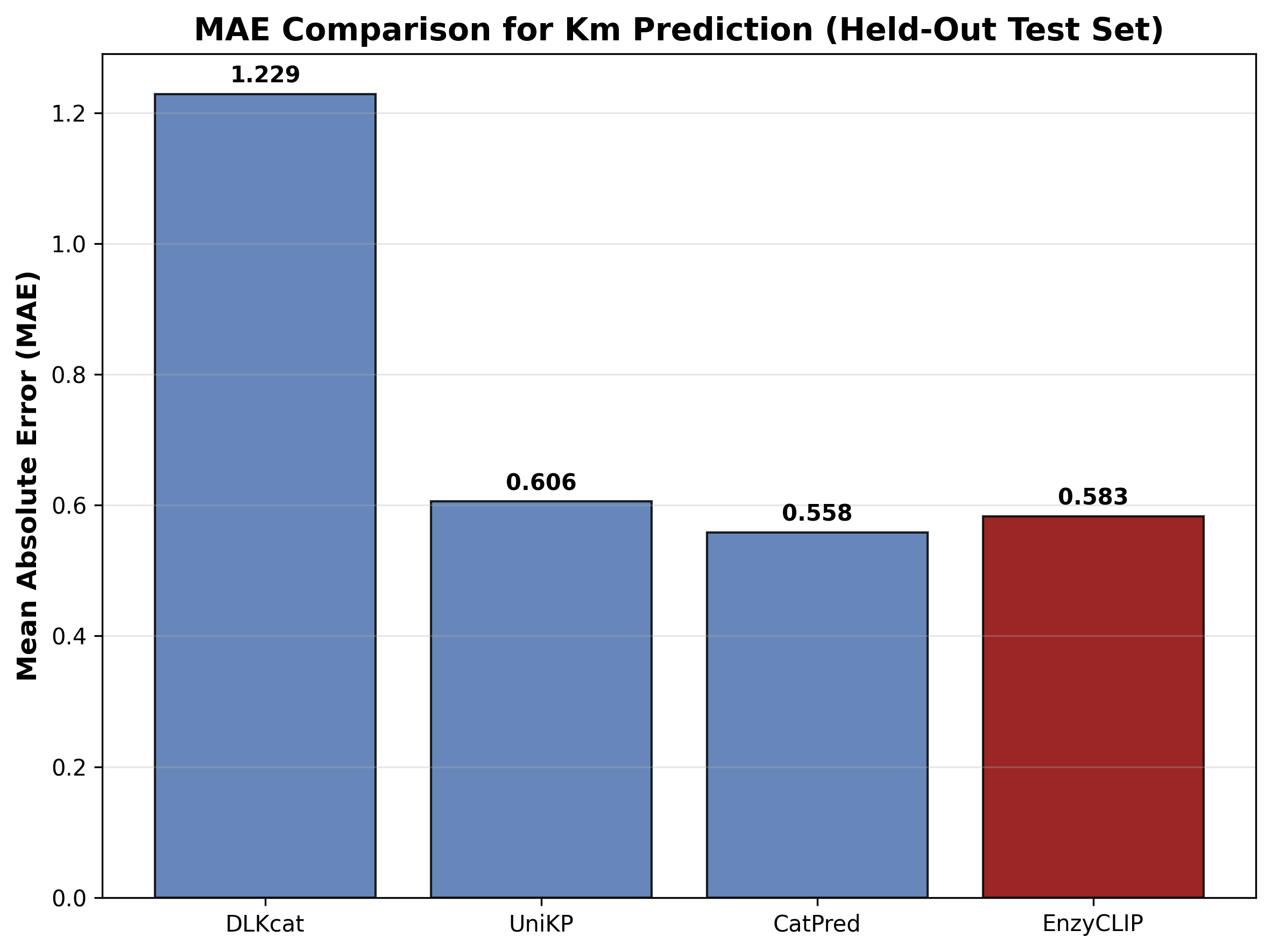}
\caption{MAE comparison across models for \(K_m\) prediction on the held-out test set. CatPred attains the lowest MAE (0.558), followed by EnzyCLIP (0.583) and UniKP (0.606), while DLKcat again exhibits high error (1.229), indicating weak predictive performance for binding affinity.}
\label{fig:km_mae_comparison}
\end{figure}

\begin{figure}[H]
\centering
\includegraphics[width=0.70\textwidth]{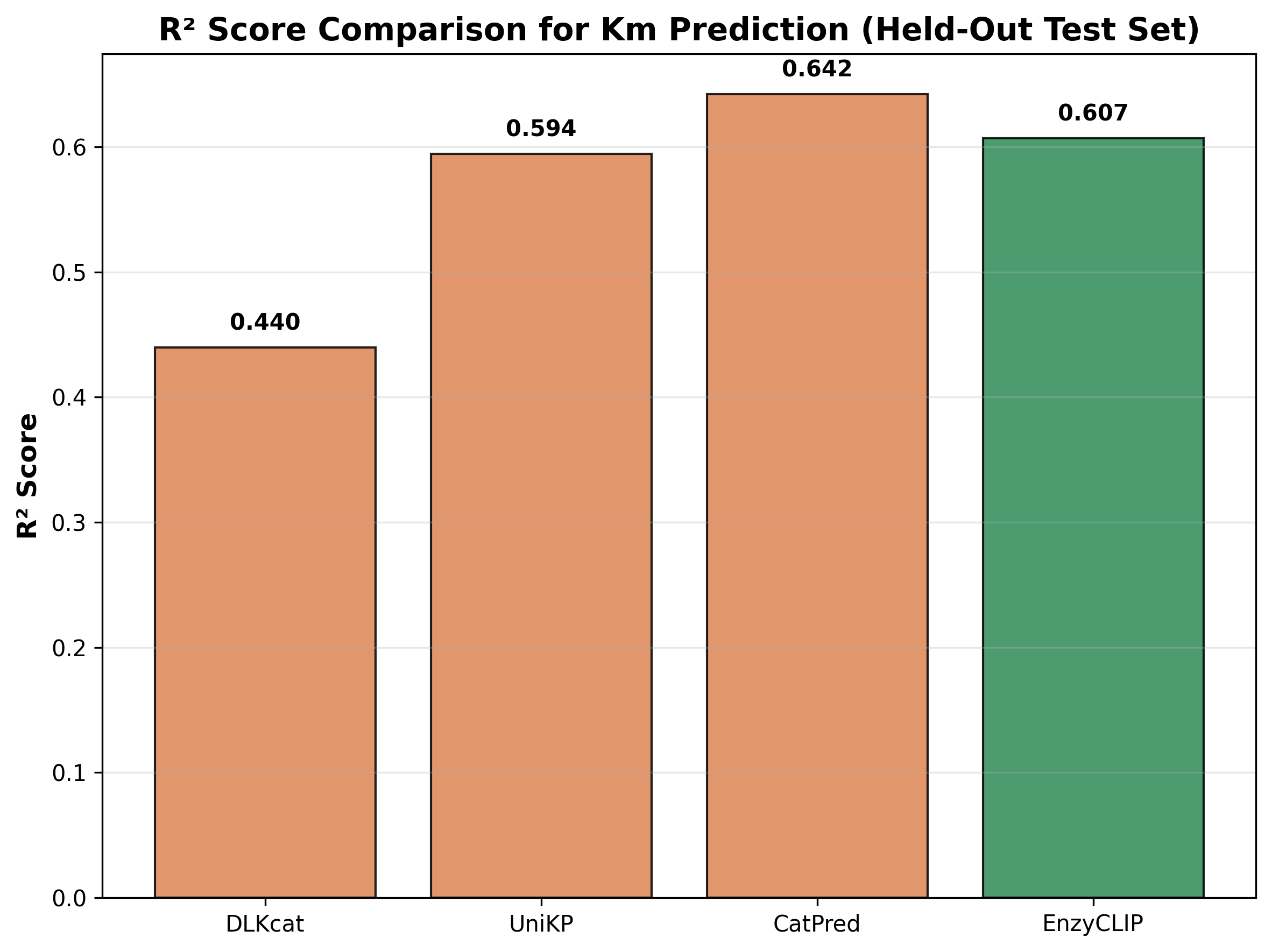}
\caption{\(R^2\) score comparison across models for \(K_m\) prediction on the held-out test set. CatPred performs best with \(R^2 = 0.642\), followed by EnzyCLIP (\(R^2 = 0.607\)) and UniKP (\(R^2 = 0.594\)), whereas DLKcat exhibits substantially lower predictive power with \(R^2 = 0.440\).}
\label{fig:km_r2_comparison}
\end{figure}

\section{Discussion}

The EnzyCLIP framework demonstrates that contrastive learning can effectively capture the relationship between enzyme sequences and substrate structures to predict kinetic parameters. The performance that the model achieves for both kinetic parameters is comparable or better than the performance of specialized methods, reaching \(R^2 = 0.607\) for \(K_m\) and \(R^2 = 0.593\) for \(K_{\text{cat}}\); this validates the approach of learning shared embedding spaces in which functional similarity is encoded as geometric proximity (Fig.~\ref{fig:kcat_baseline_evaluation}, Fig.~\ref{fig:km_baseline_plots}). While \(K_{\text{cat}}\) reflects the catalytic mechanism and \(K_m\) reflects binding affinity, both kinetic parameters have been successfully predicted with a consistent architecture, thereby illustrating the generalizability of the learned multimodal representations (Fig.~\ref{fig:kcat_plots_summarized}, Fig.~\ref{fig:km_plots_summary}).

Mechanistic insight is provided by divergent performance patterns across sequence lengths and EC classes. Degraded \(K_{\text{cat}}\) prediction for very long sequences (\(>800\) amino acids) but maintained \(K_m\) accuracy suggests that catalytic efficiency may depend on long-range allosteric effects and domain interactions that are challenging to capture from primary sequence alone (Fig.~\ref{fig:kcat_sequence_length}, Fig.~\ref{fig:km_sequence_length}). By contrast, substrate binding affinity appears more robustly predictable from local sequence features encoding active site residues (Fig.~\ref{fig:km_sequence_length}). This interpretation is consistent with the view provided by structural biology that binding pockets are created by localized sequence motifs, whereas catalytic competence often depends on the precise positioning of distant functional groups (Fig.~\ref{fig:kcat_sequence_length}, Fig.~\ref{fig:km_sequence_length}).

We observed through SHAP-based explainability analysis that some embedding dimensions are consistently driving the predictions, while different feature importance profiles emerge for \(K_{\text{cat}}\) versus \(K_m\). The dimensional specialization here illustrates that contrastive learning naturally segregates features relevant for different kinetic properties, an emergent organization not explicitly programmed into the model architecture. The high-importance dimensions identified herein allow for targeted model interpretation and hypothesis generation on sequence–function relationships.

That said, it was observed that ensemble modeling with XGBoost provided consistent improvements across both parameters and gave evidence that gradient boosting operates well on the learned representations. The magnitude of improvement, 1--5\% \(R^2\) gain, suggests that while EnzyCLIP captures the primary signal, nonlinear feature interactions remain that are better modeled by decision tree ensembles. The two-stage approach-contrastive representation learning followed by supervised ensemble regression-combines the strengths of self-supervised and supervised learning paradigms. These results position EnzyCLIP as a state-of-the-art method in enzyme kinetic prediction competitive with specialized methods, offering the advantage of a unified framework applicable for multiple kinetic parameters. The explainability via SHAP analysis allows trustworthiness and biologically insightful interpretation-a significant gap in black-box machine learning approaches in biochemical applications. Further improvements could be achieved in future work using structure information, dynamical conformational ensembles, and phylogenetic information, especially for difficult to predict multi-functional enzymes or enzymes with allosteric regulation.

\section{Conclusion}

Across all benchmark comparisons, EnzyCLIP demonstrates that a contrastive multimodal architecture can be on par or even outperform several specialized enzyme kinetics prediction models while being much lighter and easier to deploy. During the \(K_{\text{cat}}\) evaluations, EnzyCLIP achieved competitive agreement with state-of-the-art models such as CatPred and UniKP, while significantly outperforming DLKcat for both \(R^2\) and MAE (Fig.~\ref{fig:kcat_mae_comparison}, Fig.~\ref{fig:kcat_r2_comparison}). A similar trend was seen for \(K_m\), where EnzyCLIP is among the top performers and showed high predictive consistency across all metrics (Fig.~\ref {fig:km_mae_comparison}, Fig.~\ref{fig:km_r2_comparison}). This points out that, beyond its auxiliary use, contrastive representation learning can serve as a robust basis for kinetic parameter modeling.

One of the main advantages of EnzyCLIP refers to its practical usability: the framework is based on the smallest ESM-2 model (with 8M parameters), which allows very fast training with minimal computational overhead. Therefore, this approach is very reproducible, accessible for researchers without GPU clusters, and well adapted for fast experimentation. Its efficiency also allows testing new organism-specific datasets, screening tasks, or curated subsets of RENDA and SABIO-RK without the barriers typically associated with large protein language models. 

All code, training scripts, and example notebooks are openly available in a single repository to ensure full reproducibility and ease of adoption by the community:
\[
\text{\url{https://github.com/Anasazizkhan/EnzyCLIP}}
\]

In all, EnzyCLIP offers a fast, light, and extensible enzyme kinetics prediction platform that is competitive in accuracy with specialized models but offers substantial advantages in terms of scalability and reproducibility. Performance, interpretability, and accessibility make it a useful tool in computational enzymology, biocatalyst design, and modeling of large-scale biochemical data.
\appendix

\section*{Appendix}
Additional methodological details, experimental notes, or supplementary explanations can be added here if needed. Currently, no supplementary materials are included.

\bibliography{sn-bibliography}

\end{document}